\documentclass[reprint,amsmath,aps,prd,floatfix]{revtex4-1}
\pdfoutput=1
\usepackage[colorlinks, linkcolor=blue, citecolor=blue]{hyperref}

\usepackage[T1]{fontenc}
\usepackage{graphicx}

\usepackage{slashed}
\usepackage{longtable, supertabular, multirow}
\usepackage{cleveref}
\usepackage{float}

\usepackage{bm}
\newcommand{\bb}[1]{\bm{\mathrm{#1}}}

\usepackage{siunitx}
\DeclareSIUnit{\year}{yr}
\newcommand\meV{\milli\electronvolt}
\newcommand\eV{\electronvolt}
\newcommand\keV{\kilo\electronvolt}
\newcommand\MeV{\mega\electronvolt}
\newcommand\GeV{\giga\electronvolt}

\usepackage{array}
\newlength\Cwidth
\newcolumntype{C}{>{\hspace{2mm}$}p{\Cwidth}<{$\hfil}}

\newcommand{\du}{\mathrm{d}}
\newcommand{\dd}{\,\du}
\newcommand{\fd}[2]{\frac{\du #1}{\du #2}}
\newcommand{\abs}[1]{\left|#1\right|}
\newcommand{\avg}[1]{\left\langle#1\right\rangle}
\newcommand{\Neff}{N_{\mathrm{eff}}}
\newcommand{\dm}{\chi} 
\newcommand{\dmf}{\psi} 
\newcommand{\dms}{\phi} 
\newcommand{\dmmed}{\zeta} 
\newcommand{\TFO}{T_{\mathrm{FO}}}
\newcommand{\TBBN}{T_{\mathrm{BBN}}}
\newcommand{\BBNThresholdT}{\SI{1}{\MeV}}
\newcommand{\coup}{g}
\newcommand{\supp}{\Lambda_{\mathrm{EFT}}}
\newcommand{\scalaroperator}[1]{\mathcal O^{(\dms)}_{#1}}
\newcommand{\fermionoperator}[1]{\mathcal O^{(\dmf)}_{#1}}
\newcommand{\plotscale}{0.70}

\newcommand\constraintlabels{\ignorespaces
    Background contours show scattering cross section, labeled as $\log_{10}(\sigma_{\mathrm{scat}}/\SI{}{\centi\meter^2})$.
    \textit{Green, CMB:} constraint from $\Neff$.
    \textit{Orange, BBN:} solid line: constraint from light element abundances with a threshold temperature of \BBNThresholdT. Dashed line: constraint with a threshold temperature of \SI{2.3}{\MeV} (see \cref{sec:bbn}).
    \textit{Blue, RD:} constraint from relic density.
    \textit{Black, DD:} direct detection sensitivity (95\% CL) with \SI{1}{\kilo\gram.\year} exposure.
}

\begin{document}

\setcounter{tocdepth}{2}

\title{Cosmology and prospects for sub-MeV dark matter in electron recoil experiments}

\author{Benjamin V. Lehmann}
\email{blehmann@ucsc.edu}

\author{Stefano Profumo}
\email{profumo@ucsc.edu}

\affiliation{Department of Physics, 1156 High St., University of California Santa Cruz, Santa Cruz, CA 95064, USA}
\affiliation{Santa Cruz Institute for Particle Physics, 1156 High St., Santa Cruz, CA 95064, USA}

\begin{abstract}\ignorespaces
    Dark matter is poorly constrained by direct detection experiments at masses below \SI{1}{\MeV}. This is an important target for the next generation of experiments, and several methods have been proposed to probe this mass range. One class of such experiments will search for dark matter--electron recoils. However, simplified models with new light degrees of freedom coupled to electrons face significant pressure from cosmology, and the extent of these restrictions more generally is poorly understood. Here, we perform a systematic study of cosmological constraints on models with a heavy mediator in the context of an effective field theory. We include constraints from (i) disruption of primordial nucleosynthesis, (ii) overproduction of dark matter, and (iii) the effective number of neutrino species at recombination. We demonstrate the implications of our results for proposed electron recoil experiments, and highlight scenarios which may be amenable to direct detection.
\end{abstract}

\maketitle

\section{Introduction}
The identity of dark matter (DM) remains one of the most significant problems in cosmology and particle physics. Over the past few decades, experimental efforts to detect and characterize DM have been guided by the assumption that the dark species is a weakly interacting massive particle (WIMP). However, despite substantial improvements to experimental sensitivity, neither astrophysical nor terrestrially-produced DM has been definitively detected. Increasingly strong constraints have placed the WIMP paradigm under pressure \citep{Kahlhoefer:2017dnp,Agnese:2017njq,Aprile:2018dbl,Cui:2017nnn,Arcadi:2017kky}, spurring the development of new models across the mass spectrum.

In the meantime, direct searches for DM have largely targeted the weak scale. Most extant direct detection experiments are designed to detect the scattering of DM with atomic nuclei, and due to kinematic limits, they have poor sensitivity to a DM particle with mass below \SI{10}{\GeV} \citep{Akerib:2013tjd,Undagoitia2015,Tan:2016zwf}. Analyses of the phase space distribution of DM in dwarf spheroidal galaxies bound the mass of fermionic DM to $m_{\mathrm{DM}}\gtrsim\SI{1}{\kilo\electronvolt}$ regardless of the production mechanism \cite{Boyarsky:2008ju}, and the Lyman-$\alpha$ forest imposes a comparable constraint on thermal relic DM of any kind \citep{Viel:2005qj}. But beyond these bounds, DM models with mass between \SI{1}{\keV} and \SI{10}{\GeV} are poorly constrained. Several well-motivated scenarios \citep[e.g. asymmetric DM,][]{Zurek:2013wia} naturally feature masses between \SI{1}{\keV} and \SI{10}{\GeV}, making this range an appealing target for future direct detection experiments \cite{Caputo:2019cit}.

This has driven much interest in novel detection methods suited to light DM particles, and several such experiments have been proposed in the last few years \citep{Essig:2011nj,Essig:2015cda,Hochberg2016a,Hochberg2016b,Hochberg2016c,Derenzo:2016fse,Hochberg2016d,Knapen2016,Hochberg:2017wce,Kurinsky:2019pgb,Abdelhameed:2019hmk} (see sections IV--V of \cite{Battaglieri:2017aum} for a review). These experiments are designed to be sensitive to the very small recoil energies characteristic of the scattering of light particles, and as such, many are designed to search for the scattering of DM with electrons instead of nuclei, a strategy first detailed in \cite{Essig:2011nj}. Several experiments now constrain DM--electron scattering at masses as low as $\sim\SI{1}{\MeV}$ \cite{Graham:2012su,Essig:2017kqs,Agnes:2018oej,Agnese:2018col,Abramoff:2019dfb,Aguilar-Arevalo:2019wdi}. The more recent proposal of \cite{Hochberg2016a}, based on electrons in aluminum superconductors, is sensitive to deposited energies of order \SI{1}{\meV}, allowing for the detection of particles as light as \SI{1}{\keV}.

However, although the most generic astrophysical constraints do not restrict DM at masses between \SI{1}{\keV} and \SI{10}{\GeV}, it is well known that particular models can be constrained by cosmological observables, especially for masses below $\SI{1}{\MeV}$ \cite{Green2017}. In particular, light DM interacting with electrons risks running afoul of the following restrictions:
\begin{itemize}
\item The DM must not significantly alter successful predictions of the ratios of light elemental abundances produced in big bang nucleosynthesis (BBN) \citep{Kolb:1986nf,Serpico:2004nm,Jedamzik:2009uy};
\item To accord with measurements of the effective number of neutrino species ($\Neff$), the thermal history of the DM species must not significantly alter the temperature ratio of photons and neutrinos at recombination \citep{Boehm2013};
\item While a single species of DM particle may not account for the entirety of the present-day DM density, no species may be produced with an abundance exceeding that threshold.
\end{itemize}
In each case, such cosmological constraints bound the couplings between new species and Standard Model (SM) particles, which also determine the event rates in direct detection experiments. Thus, in a given model, the cosmological effects of light DM can be related to the direct detection cross section. Given an experimental proposal and a DM model, one can then determine the extent of the parameter space accessible to the experiment and consistent with cosmology. Such an approach has been applied to electron recoil experiments by~\cite{Knapen:2017xzo} for a class of simplified models, and more recently in a variety of model-dependent instances \cite{Berlin:2017ftj,Krnjaic:2017tio,Depta:2019lbe,Berlin:2019pbq,Digman:2019wdm,Bondarenko:2019vrb,Sabti:2019mhn,Chang:2019xva,Fiaschi:2019evv}.

In this work, we show that cosmological constraints on a new light (sub-MeV) species interacting with electrons can be greatly generalized with a small number of assumptions. Assuming a heavy mediator between DM and the SM, we study the cosmological implications of a light DM species in an effective field theory (EFT), and use the same EFT to evaluate direct detection prospects. We thus obtain model-independent cosmological limits on the scattering cross section of DM with electrons in an actual experiment. The model-independent methodology is similar in spirit to \cite{Bertuzzo:2017lwt,Choudhury:2019tss,Caputo:2019ywq}, but applied to directly connect cosmological constraints and detection prospects in the sub-MeV regime.

This paper is organized as follows. In \cref{sec:EFT}, we describe our EFT framework for modeling light DM coupled to electrons. In \cref{sec:thermal-history}, we derive model-independent cosmological constraints on the DM species. In \cref{sec:constraints}, we evaluate the DM--SM scattering cross section in our EFT, and compare cosmological bounds with prospects in a fiducial experiment. Finally, we discuss implications for direct detection experiments in \cref{sec:discussion}. A complete set of constraints and tables of cross sections are placed after the end of the text.

Throughout this work, we denote a scalar DM field by $\dms$ and a fermionic DM field by $\dmf$. When speaking about the DM species generally, without specifying its spin, we will denote it with $\dm$.

\section{Effective interactions of sub-MeV dark matter}
\label{sec:EFT}
In this section, we build a theoretical framework to study the effective interactions of sub-MeV DM of spin 0 or $\frac12$. We study DM candidates that are singlets under the SM gauge groups, and we consider both scalar and fermionic DM. We first specify the working assumptions of our EFT framework, and we thereafter develop the scalar and fermion cases separately.

\subsection{The EFT framework}
We assume that DM is dominated by a single particle species with a mass below \SI{1}{\MeV}. The MeV scale is cosmologically significant as the scale of big-bang nucleosynthesis (BBN). The DM annihilation and scattering processes that we consider in this work always involve energy exchanges well below this scale, whether they take place in the early universe or in a laboratory today. Thus, this situation lends itself well to an effective low-energy description with an EFT that has a cutoff of order \SI{10}{\MeV}. In general, the EFT can be valid up to higher scales, but since cosmological history is poorly constrained at temperatures above a few MeV, we only apply the EFT at or below this scale.

At energies well below the MeV scale, the only dynamical SM degrees of freedoms are electrons and positrons ($e^\pm$), neutrinos ($\nu$), and photons ($\gamma$). We assume further that there is no additional light degree of freedom besides the DM particle: all remaining new physics is presumed to lie well above the MeV scale, including any mediators between DM and SM particles. Physics at sub-MeV scales is thus well described by an EFT in which only $e^\pm$, $\nu$, $\gamma$, and the DM $\dm$ are dynamical degrees of freedom. This is the theoretical framework we employ for our analysis.

\begin{figure*}\centering
\includegraphics[width=0.6\linewidth]{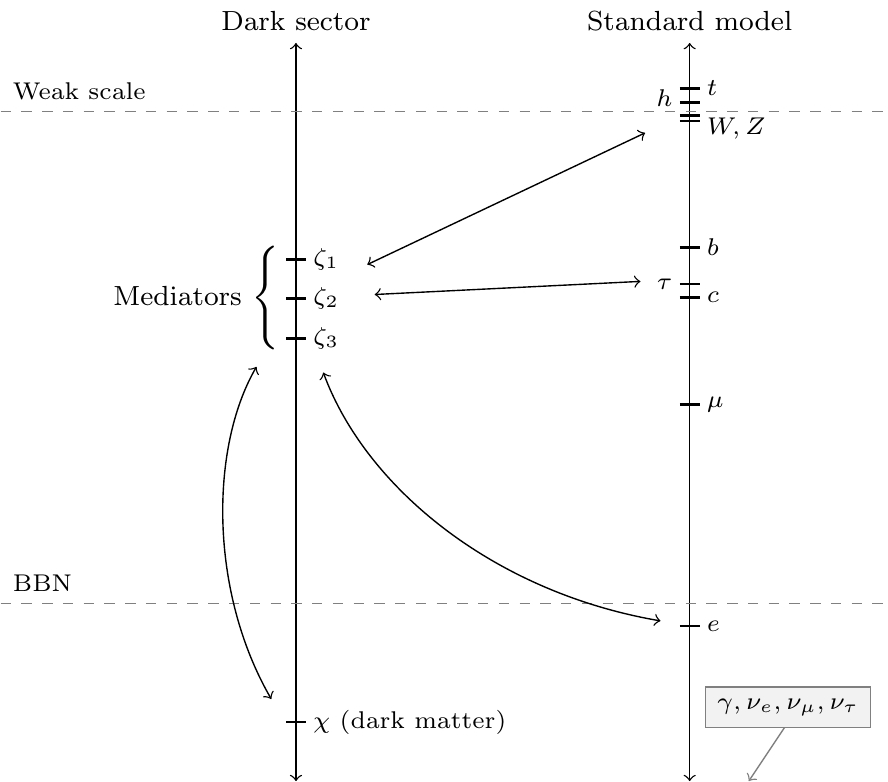}
\caption{Schematic description of a UV completion of our effective theory. The vertical direction on the diagram corresponds to the mass scale. Arrows denote renormalizable couplings. Note that there is no renormalizable interaction between the DM and SM fields. The line labeled ``BBN'' corresponds to the scale of big bang nucleosynthesis, $T\sim\SI{1}{\MeV}$. Our results are unchanged if $m_{\zeta_i}>m_{\mathrm{weak}}$.}
\label{fig:sectors}
\end{figure*}

Before presenting the EFT in more detail, it is instructive to take a step back and discuss the conceptual starting point of our work: a renormalizable theory with DM as well as mediator fields in the spectrum. The EFT language powerfully encodes the many UV-complete realizations which give the same low energy physics. We make three additional assumptions about the UV-complete theory, described below and graphically summarized in \cref{fig:sectors}:
\begin{enumerate}
\item The DM is stablilized by a $Z_2$ symmetry and is thus absolutely stable.
\item The couplings between mediators and SM fields respect electroweak gauge invariance, in the sense that the $\dm$--$e_L$ coupling is equal to the $\dm$--$\nu$ coupling. We make this assumption to clarify the impact of the DM species on $\Neff$, as discussed in the next section. It does not influence the other constraints.
\item DM couples to the visible sector via mediator fields $\zeta_i$, with masses satisfying $\TBBN\ll m_{\zeta_i}$.
\end{enumerate}

When writing our EFT Lagrangian, it is convenient to take $m_{\zeta_i}\ll m_{\mathrm{weak}} \simeq \SI{100}{\GeV}$, so that weak-scale degrees of freedom in the SM can be integrated out before the mediators. It is then possible to define an intermediate EFT with weak scale particles integrated out and mediators in the spectrum. However, our results do not depend on this assumption---it simply clarifies how we should write the low-energy Lagrangian to accommodate lower mediator masses.

Ultimately, our EFT will contain a mass scale $\supp$ which is related to the mediator masses, and each operator will appear with a coupling (Wilson coefficient) $\coup$. We ensure that we remain in the regime of validity of the EFT by enforcing $\supp\gg\TBBN$, so it is convenient to assume that $\coup\sim\mathcal O(1)$ and take $\supp$ to be the free parameter in our analysis. Small deviations of $\coup$ from unity can then be absorbed by rescaling $\supp$. But if $\coup$ is not $\mathcal O(1)$ in a typical UV completion, and $\supp$ is not many orders of magnitude larger than $\TBBN$, we have reason for caution: rescaling $\supp$ to absorb a very small $\coup$ could violate the requirement that $\supp\gg\TBBN$. Thus, when the scale of the DM--SM interaction is smaller, it is important to separate $\coup$ from any non-$\mathcal O(1)$ coupling typical of UV completions. An intermediate EFT lying below the weak scale guides our expectations for the size of the coupling in the effective theory after integrating out the mediators.

In particular, if a scalar $\zeta$ mediates the DM--SM interaction, it is easy to generate a factor of the electron Yukawa coupling $y_e$. Coupling $\zeta$ to the lepton doublet $L$ without breaking gauge invariance involves interaction terms of the form
\begin{multline}
    \mathcal L_{\mathrm{UV}}\supset
        M_1\zeta\dms^\dagger\dms
        + M_2\zeta H^\dagger H \\
        + \zeta^\dagger\zeta H^\dagger H
        + y_e\bar LHe_R+\mathrm{c.c.}
\end{multline}
Thus, after EWSB, $\zeta$ mixes with the Higgs boson $h$. To construct an EFT from the Lagrangian in the broken phase, we must integrate out the mass eigenstates corresponding to $(\zeta, h)$, which will always produce a factor of $y_e$ in addition to the inverse of the mediator mass scale.

Such a factor of $y_e$ in the EFT is also expected on general grounds if minimal flavor violation is assumed, regardless of the nature of the mediator. However, in general, one can also write UV completions which do not generate a factor of $y_e$, e.g. by employing a vector mediator. Still other UV completions can be constructed to introduce other small coefficients besides $y_e$ in the EFT. When we tabulate the EFT operators, to facilitate comparison with arbitrary UV completions, we do not normalize the operators with such any such factor. However, since a factor of $y_e$ is well-motivated, we will give our results in a format that shows constraints both with and without a factor of $y_e$.

Finally, note that we ignore any renormalizable couplings between the DM and SM fields, assuming that all interactions are encoded in the EFT. Notice that no such operators exist in the fermionic case under our assumptions, since we take the DM to be a SM singlet, and the $Z_2$ symmetry forbids the lepton portal operator $\dms L H$. In the scalar case, on the other hand, this is something we impose. However, as we will discuss shortly, this assumption has no consequences for the results of our analysis.

At energies at or below the scale of BBN, the effective Lagrangian schematically reads
\begin{equation}
\label{eq:eft-lagrangian}
\mathcal{L}_{\mathrm{EFT}} =
    \mathcal{L}_{\mathrm{SM}} + \mathcal{L}_{\mathrm{DM}} + 
    \sum_{d > 4, \alpha}
        \frac{c_{\alpha}}{\supp^{d - 4}} \mathcal{O}_{\alpha}
    .
\end{equation}
Here $\supp$ is the mass scale associated with the EFT, which reflects the scale of the heavy degrees of freedom in the theory; $\mathcal L_{\mathrm{SM}}$ is the SM Lagrangian with only the $e^\pm$, $\nu$, and $\gamma$ fields; and $\mathcal L_{\mathrm{DM}}$ is the DM free theory contribution. The form of $\mathcal L_{\mathrm{DM}}$ depends on whether the DM is a scalar $\dms$ or a fermion $\dmf$. If the DM is a scalar, then
\begin{equation}
    \mathcal{L}_{\mathrm{DM}} = \mathcal{L}_{\dms} = \begin{cases}
        \frac{1}{2} \partial^\mu \dms \partial_\mu \dms
            - \frac{1}{2} m_\dms^2 \dms^2 
            & \text{real scalar}
        \\
        (\partial^\mu \dms)^\dag (\partial_\mu \dms)
            - m_\dms^2 \dms^\dag \dms
            & \text{complex scalar,}
    \end{cases}
\end{equation}
and if the DM is a fermion, then
\begin{equation}
    \mathcal{L}_{\rm DM} = \mathcal{L}_{\dmf} = \begin{cases}
        \frac{1}{2} \bar{\dmf} i \slashed{\partial} \dmf
            - \frac{1}{2} m_\dmf \bar{\dmf} \dmf
            & \text{Majorana fermion}
        \\
        \bar{\dmf} i \slashed{\partial} \dmf
            - m_\dmf \bar{\dmf} \dmf
            & \text{Dirac fermion.}
    \end{cases}
\end{equation}
The remaining (infinite) sum over the higher-dimensional operators in \cref{eq:eft-lagrangian} accounts for the effective interactions between DM and SM fields. In our analysis, we will retain terms up to dimension 6.

In the following subsection, we parametrize the interactions between DM and electrons. All operators consistent with a $Z_2$ symmetry have the schematic form 
\begin{equation}
\mathcal{O}^{(\dm)} \propto B_I(\dm) \; \bar{e} \, \Gamma^I e,
\label{eq:bilinear-general}
\end{equation}
where the function $B_I(\dm)$ contains an even number of DM fields, and $I$ denotes a set of Lorentz indices. We will eventually truncate all operators beyond dimension 6, so for our purposes, $B_I(\dm)$ always contains two DM fields. This DM bilinear is multiplied by an electron bilinear, for which the independent Dirac structures can be fully enumerated:
\begin{equation}
    \label{eq:dirac-bilinears}
    \Gamma^I \in \operatorname{span}\left\{
        1 , \,  i\gamma^5 , \, \gamma^\mu ,  \,
        \gamma^\mu \gamma^5 , \,  \sigma^{\mu\nu}
    \right\}.
\end{equation}
If the electron bilinear is not a Lorentz scalar, the contraction of its free Lorentz indices with the ones of the DM bilinear ensures that the full operator in \cref{eq:bilinear-general} is a Lorentz invariant. We now discuss the allowed operators for scalar and fermion DM.

\subsection{EFT for scalar DM}
\label{sec:eft-scalar}
To describe our EFT for scalar DM, we must enumerate all operators of the form
\begin{equation}
    \mathcal{O}^{(\dms)} \propto B_I(\dms) \; \bar{e} \, \Gamma^I e
\end{equation}
up to some mass dimension. Note that $\dms$ carries no Lorentz indices or spinor indices. Thus, if the index set $I$ carried by the electron bilinear is non-empty, the only option is to insert derivatives in the scalar bilinear so that all indices are contracted.

A classification of all possible cases is provided in~\cref{tab:eft-scalar}. Of the four resulting operators, two are dimension-5, while the other two include a derivative and are dimension-6. We use the notation 
\begin{equation}
    \dms^\dagger\overset\leftrightarrow\partial_\mu\dms
        \equiv
    \dms^\dagger\partial_\mu\dms-(\partial_\mu\dms^\dagger)\dms.
\end{equation}
Note that we omit the operator $\left(\partial_\mu\dms^\dagger\,\dms+\dms^\dagger\,\partial_\mu\dms\right)\bar e\gamma^\mu e$, since it vanishes under integration by parts and application of the equation of motion:
\begin{multline}
    \int\du^4x\,\left(\partial_\mu\dms^\dagger\,\dms
        + \dms^\dagger\,\partial_\mu\dms\right)\bar e\gamma^\mu e\\
        =
    -\int\du^4x\,\dms^\dagger\dms\,\partial_\mu\left(
        \bar e\gamma^\mu e
    \right)=0.
\end{multline}
Similarly, the operator $\left(\partial_\mu\dms^\dagger\,\dms+\dms^\dagger\,\partial_\mu\dms\right)\bar e\gamma^\mu\gamma^5 e$ is redundant: integrating by parts again, we obtain
\begin{align}
	\int\du^4x\,\partial_\mu\bigl(\dms^\dagger\dms\bigr)
        \bar e\gamma^\mu \gamma^5 e
	&=
		-\int\du^4x\,\dms^\dagger\dms\,\partial_\mu\left(
            \bar e\gamma^\mu \gamma^5 e\right)\\
	&= -2im_e\int\du^4x\,\dms^\dagger\dms\,\bar e\gamma^5e.
\end{align}
The resulting integrand is proportional to $\scalaroperator{P}$, one of the other operators in our basis. Moreover, this contribution is dimension-6 while $\scalaroperator{P}$ is dimension-5, so it is suppressed in the Lagrangian with an additional factor of $\supp^{-1}$.

In some cases, renormalizable operators are allowed, and might appear in addition to the effective operators discussed above. For instance, in the context of a Higgs portal model \citep[see e.g.][]{Arcadi:2019lka} the operator $\dms^\dagger \dms \, H^\dag H$ is allowed without affecting DM stability. After electroweak symmetry breaking (EWSB), this operator produces  a cubic coupling $\dms^\dagger \dms \, v \, h$. Integrating out the SM Higgs boson generates an effective operator proportional to $\mathcal{O}^{(\dms)}_S$. Thus, adding renormalizable couplings does not introduce any new physical effects in our analysis. The only effect is to add a correction to the Wilson coefficient of a single operator, with a size typically smaller than the values we consider in our analysis.

At a qualitative level, we can guess at the relative prospects for direct detection in the case of each operator in \cref{tab:eft-scalar}. The operator $\scalaroperator{S}$ is easily generated by integrating out a scalar mediator, so we can expect that the relative strength of constraints and detection prospects for this operator will be comparable to results found in the context of simplified models with a scalar mediator \cite{Knapen:2017xzo}. Unlike $\scalaroperator{S}$, the other operators for scalar DM are suppressed by their momemtum dependence in the non-relativistic limit, relevant for scattering. Each of these operators vanishes as the velocity and momentum transfer are taken to zero. Thus, for scalar dark matter, we expect from the outset that none of our operators will improve on the detection prospects of a simplified model with a scalar mediator, and we will indeed confirm these suspicions in the following sections.

With the effective operators in the scalar case now enumerated, we can consider annihilation and scattering processes for each one. Matrix elements for $2\to2$ annihilation and scattering are given in \cref{tab:scalar-squared-matrix-elements}. The corresponding cross sections are given in~\cref{tab:scalar-annihilation-cross-sections,tab:scalar-scattering-cross-sections}.

\setlength\Cwidth{5cm}
\begin{table}
	\begin{equation*}
		\def\arraystretch{1.5}
		\begin{array}{| c | C | c |}
			\hline
			\text{Symbol} & \text{Operator} & \text{Real case}
			\\\hline
			\scalaroperator{S} &
				\coup\supp^{-1}\dms^\dagger\dms\,\bar ee &
				\text{Yes}
			\\
			\scalaroperator{P} &
				i\coup\supp^{-1}\dms^\dagger\dms\,\bar e \gamma^5e &
				\text{Yes}
			\\
			\scalaroperator{V} &
				i\coup\supp^{-2}\dms^\dagger\overset\leftrightarrow\partial_\mu\dms\,\bar e\gamma^\mu e &
				\text{No}
			\\
			\scalaroperator{A} &
				i\coup\supp^{-2}\dms^\dagger\overset\leftrightarrow\partial_\mu\dms\,\bar e\gamma^\mu \gamma^5 e &
				\text{No}
			\\\hline
		\end{array}
	\end{equation*}
	\caption{Operators coupling the electron to a dark scalar $\dms$. The third column indicates whether or not the operator survives when $\dms$ is taken to be a real scalar.}
	\label{tab:eft-scalar}
\end{table}

\subsection{EFT for fermion DM}
If the DM is a fermion $\dmf$, the structure of the EFT is similar to the scalar case. We again have a set of operators which are products of an electron bilinear and a $\dmf$ bilinear. Using generalized Fierz identities, it can be shown that operators of the form $(\bar\dmf\mathcal O_1 e)(\bar e\mathcal O_2\dmf)$ are redundant, in that they can be written as linear combinations of operators of the form $(\bar\dmf\mathcal O^\prime_1\dmf)(\bar e\mathcal O^\prime_2e)$ \citep{Nieves2003}. Thus, we can construct a complete basis of effective operators by enumerating the possible insertions $\mathcal O^\prime_1$ and $\mathcal O^\prime_2$. All of the electron bilinears from the scalar case appear here as well, and most of the possible $\dmf$ bilinears are obtained from these by making the replacement $e\to\dmf$.

In addition to these bilinears, we can form a spin-2 current at dimension 6, e.g. of the form $\bar\dmf\sigma_{\mu\nu}\dmf$. Since $\sigma_{\mu\nu}$ is antisymmetric, the other bilinear must not be symmetric in its Lorentz indices, so it must contain another insertion of $\sigma_{\mu\nu}$. Thus, such an operator has the general form $W_{\mu\nu\alpha\beta}\bar\dmf\sigma^{\mu\nu}\dmf\bar e\sigma^{\alpha\beta}e$. At dimension 6, the indices of $W_{\mu\nu\alpha\beta}$ can come only from two factors of the metric or one factor of the Levi-Civita symbol $\varepsilon$. In the former case, again due to antisymmetry of $\sigma^{\mu\nu}$, the only nontrivial contraction is 
\begin{equation}
    g_{\mu\alpha}g_{\nu\beta}\bar\dmf\sigma^{\mu\nu}\dmf\bar e\sigma^{\alpha\beta}e.
\end{equation}
If $W$ is instead formed from the Levi-Civita symbol, then the operator has the form $\varepsilon_{\rho_1\rho_2\rho_3\rho_4}\bar\dmf\sigma^{\mu\nu}\dmf\bar e\sigma^{\alpha\beta}e$, where $(\rho_1,\rho_2,\rho_3,\rho_4)$ is a permutation of $(\mu,\nu,\alpha,\beta)$. Up to an overall sign, the indices $\rho_i$ can be rearranged into the latter order, so all such operators are proportional to
\begin{equation}
    \bar\dmf\sigma^{\mu\nu}\dmf\bar e\left(\varepsilon_{\mu\nu\alpha\beta}\sigma^{\alpha\beta}\right)e.
\end{equation}
But $\varepsilon_{\mu\nu\alpha\beta}\sigma^{\alpha\beta}=-2i\sigma_{\mu\nu}\gamma^5$, so if we simply add $i\sigma_{\mu\nu}\gamma^5$ to our list of insertions, we can assume that $W_{\mu\nu\alpha\beta}$ is a product of metric tensors. (We retain the factor of $i$ to preserve Hermiticity.) Further, the argument above demonstrates that it is sufficient to place this insertion in only one of the two bilinears: the operator formed by inserting $i\sigma_{\mu\nu}\gamma^5$ in both bilinears is redundant. We choose to place this insertion in the electron bilinear.

The complete list of operators for fermionic DM is shown in \cref{tab:eft-fermion}. Matrix elements for $2\to2$ annihilation and scattering are given in \cref{tab:dirac-squared-matrix-elements}. The corresponding cross sections are given in~\cref{tab:dirac-annihilation-cross-sections,tab:dirac-scattering-cross-sections}.

As in the scalar case, we estimate relative prospects for direct detection among the operators in \cref{tab:eft-fermion}. The operator $\fermionoperator{SS}$, like $\scalaroperator{S}$, is naturally generated by simplified models with a scalar mediator. While many of the other operators are momentum-suppressed in the non-relativistic limit, as in the case of scalar DM, the operators $\fermionoperator{VV}$, $\fermionoperator{AA}$, and $\fermionoperator{TT}$ are not. These operators may be expected to compete with or exceed the detection prospects associated with $\fermionoperator{SS}$, an expectation that we will confirm in our analysis.

\begin{table*}
	\begin{equation*}
		\def\arraystretch{1.5}
		\setlength\Cwidth{4.1cm}
		\begin{array}{| @{\hspace{2mm}}c@{\hspace{2mm}} | C | c || @{\hspace{2mm}}c@{\hspace{2mm}} | C | c |}
			\hline
			\text{Symbol} & \text{Operator} & \quad\text{Maj.}\quad\  & \text{Symbol} & \text{Operator} & \quad\text{Maj.}\quad\ 
			\\\hline
			\fermionoperator{SS} &
				\coup\supp^{-2}\bar\dmf\dmf\,\bar e e &
				\multirow{2}{*}{\text{Yes}}
			&
			\fermionoperator{PS} &
				i\coup\supp^{-2}\bar\dmf \gamma^5\dmf\,\bar e e &
				\multirow{2}{*}{\text{Yes}}
			\\
			\fermionoperator{SP} &
				i\coup\supp^{-2}\bar\dmf \dmf\,\bar e \gamma^5 e &
			&
			\fermionoperator{PP} &
				\coup\supp^{-2}\bar\dmf \gamma^5\dmf\,\bar e\gamma^5 e &
			\\
			\hline
			\fermionoperator{VV} &
				\coup\supp^{-2}\bar\dmf\gamma_\mu\dmf\,\bar e\gamma^\mu e &
				\multirow{2}{*}{\text{No}}
			&
			\fermionoperator{AV} &
				\coup\supp^{-2}\bar\dmf\gamma_\mu \gamma^5\dmf\,\bar e\gamma^\mu e &
				\multirow{2}{*}{\text{Yes}}
			\\
			\fermionoperator{VA} &
				\coup\supp^{-2}\bar\dmf\gamma_\mu\dmf\,\bar e\gamma^\mu \gamma^5 e &
			&
			\fermionoperator{AA} &
				\coup\supp^{-2}\bar\dmf\gamma_\mu \gamma^5\dmf\,\bar e\gamma^\mu \gamma^5 e &
			\\
			\hline
			\fermionoperator{TT} &
				\frac12\coup\supp^{-2}\bar\dmf\sigma_{\mu\nu}\dmf\,\bar e\sigma^{\mu\nu} e &
				\text{No}
			&
			\fermionoperator{T \tilde T} &
				\frac i2\coup\supp^{-2}\bar\dmf\sigma_{\mu\nu}\dmf\,\bar e \sigma^{\mu\nu}\gamma^5 e &
				\text{No}
			\\[0.08cm]\hline
		\end{array}
	\end{equation*}
	\caption{Operators coupling the electron to a dark fermion $\dmf$. The third column in each half of the table indicates whether or not the operator survives when $\dmf$ is taken to be a Majorana fermion.}
	\label{tab:eft-fermion}
\end{table*}

\section{Cosmological constraints}
\label{sec:thermal-history}
Cosmological constraints on DM are typically model-dependent. However, the broad class of models which we consider admits only a very restricted set of thermal histories for the DM species, which allows us to derive general cosmological constraints in the context of our EFT.

We divide the thermal histories into two cases: either the DM is in thermal equilibrium with the SM at high temperatures, and freezes out below some temperature; or it never attains thermal equilibrium, and the abundance is instead set non-thermally. It is possible that the dark species only enters equilibrium at late times, but this scenario mirrors the thermal freeze-out case in almost every respect.

In the freeze-out scenario, two constraints are particularly robust: first, if the DM is thermalized and relativistic during the epoch of big bang nucleosynthesis (BBN), its effect on the Hubble parameter is generally sufficient to perturb light elemental abundances \citep{Jedamzik:2009uy}.
Second, if at some temperature the DM is in thermal equilibrium with electrons and not with neutrinos, or vice versa, then entropy can be transferred from the DM to neutrinos alone or to electrons and photons alone. This changes the temperature ratio of the two thermal baths, which modifies the effective number of neutrino species, $\Neff$, as determined from the cosmic microwave background (CMB) \citep{Boehm2013}.

Finally, in the case of out-of-equilibrium (non-thermal) production, the DM never attains thermal equilibrium, and so may evade these two constraints. However, if the coupling to electrons is too large, DM will be overproduced even under the most generous assumptions.

Note that new light species are also subject to constraints from energy loss in stars and supernovae \cite{Raffelt:1996wa}. However, these constraints rely on complicated microphysical inputs that must be computed in detail for each model. Moreover, supernova temperatures lie up to an order of magnitude above the scale of BBN, requiring our effective theory to be valid at higher energies. Thus, we do not evaluate these constraints explicitly, but simplistic estimates suggest that they are at best comparable in strength to our cosmological constraints over the mass range of interest.

We now examine each of our constraints in more detail.

\subsection{Freeze-out and primordial nucleosynthesis}
\label{sec:bbn}
Light element abundances today are a sensitive probe of cosmology at scales near $\SI{1}{\MeV}$. If an additional light species is assumed to be in thermal equilibrium at these scales, the standard predictions of big bang nucleosynthesis (BBN) are modified, with observable consequences. Since thermal equilibrium in turn depends on DM interactions, light element abundances translate to stringent constraints on the interaction rates.

In a broad class of models, the DM species is in thermal equilibrium with the SM bath at high temperatures, and eventually drops out of equilibrium below some freeze-out temperature, $\TFO$. In our framework, freeze-out is a generic requirement of any scenario in which DM is in thermal equilibrium with electrons at temperatures $T\lesssim\BBNThresholdT$, since the EFT is valid in this regime.

If the DM species freezes out during or after BBN, and the DM species is in equilibrium at higher temperatures, then the predictions of light element abundances are generally perturbed to a degree incompatible with their measured values \cite{Kolb:1986nf,Serpico:2004nm,Cyburt:2004yc,Jedamzik:2009uy}. The ratios of these abundances are set by the temperatures at which interconversion processes freeze out, which depend in turn on the Hubble parameter $H$. Since $H$ is sensitive to the energy density, adding a new species that stays in equilibrium and remains relativistic for much of the epoch of BBN has a significant impact on the produced light element abundances. Note that in a small range of our parameter space, equilibrium during BBN is consistent with observables if the dark species \emph{enters} equilibrium at a specific time during BBN \cite{Berlin:2019pbq}. This is a very narrow exception to our framework, so we neglect it for the remainder of this work.

The temperature at which freeze-out occurs is fixed by the DM mass and the couplings. The prospect of experimental detection by any particular apparatus places a lower bound on the scattering cross section $\dm e^-\to\dm e^-$. However, for a given interaction, the scattering cross section is directly related to the annihilation cross section $\dm\dm\to e^+ e^-$ which regulates the thermodynamics of the DM species in the early universe. A lower bound on the scattering cross section thus corresponds to a lower bound on the annihilation cross section, which translates to an upper bound on the freeze-out temperature.

For our purposes, we will only consider a model to be ruled out by light element abundances if it predicts that DM is in equilibrium at $T=\BBNThresholdT$. This choice of threshold temperature is slightly different from some other treatments of BBN constraints in the literature. In particular, \cite{Boehm2013} find that sub-MeV DM is generally ruled out by elemental abundances if the DM is in equilibrium after neutrinos decouple at \SI{2.3}{\MeV}. However, these constraints assume that the DM is in equilibrium with only one of electrons and neutrinos, and not both, so that the temperature ratio $T_\nu / T_\gamma$ is modified. We will discuss this scenario in detail in the following section, but for the moment, we note that our EFT accommodates equilibrium with both electrons and neutrinos, with decouplings taking place at different temperatures. In such situations, constraints from $T_\nu / T_\gamma$ can potentially be relaxed in some areas of the parameter space. Thus, there is not necessarily any connection between neutrino decoupling and BBN constraints in our model.

Given more detailed information about the dark sector and its couplings to the SM, it is possible that BBN could place constraints on DM which decouples at even higher temperatures. Between $T\sim\SI{10}{\MeV}$ and $T=\SI{1}{\MeV}$, no SM species become non-relativistic, so the SM bath is not heated relative to a decoupled dark sector. Thus, even if the DM decouples from the SM bath at \SI{10}{\MeV} or above, it is possible that $T_\dm=T_\gamma$ during BBN, in which case sub-MeV DM will typically disrupt BBN. Additionally, if DM is in equilibrium with only one of neutrinos and electrons after neutrino decoupling takes place, then the constraints of \cite{Boehm2013} do apply.

We wish to place conservative constraints that are independent of these details, and also independent of cosmological modifications at $T\gg\SI{1}{\MeV}$ that might occur outside the context of our DM model. We regard \BBNThresholdT~as a reasonable fiducial threshold for assessing BBN constraints. However, while it is possible to avoid the constraints of \cite{Boehm2013} in our model, this takes additional tuning. Thus, we will give two versions of the BBN constraint: one with a threshold of \BBNThresholdT, and another with a threshold of \SI{2.3}{\MeV}, corresponding to the constraint of \cite{Boehm2013}. This also serves to demonstrate the sensitivity of our constraints to higher thresholds.

The freeze-out temperature and relic density for a given model are found by solving the Boltzmann equation in a relatively simple incarnation. In our framework, we have only a single DM species $\dm$ which interacts with electrons exclusively through $2\to2$ processes. For this case, using Maxwell--Boltzmann statistics, the Boltzmann equation takes the form
\begin{equation}\label{eq:boltzmann-abundances}
    \frac{x}{Y_{\mathrm{eq}}}\fd Yx
    = -\frac{
        n_{\mathrm{eq}}(x)\avg{
            \sigma\abs v
        }(x)}{H(x)}
        \left(\left(\frac{Y(x)}{Y_{\mathrm{eq}}(x)}\right)^2-1\right),
\end{equation}
where $x\equiv m_\dm/T$ parametrizes cosmic time; $\sigma$ is the cross section for $\bar\chi\chi\to e^+e^-$; $Y\equiv n/s$ is the abundance of $\dm$, where $n$ is the number density and $s$ the entropy density of $\dm$; and $Y_{\mathrm{eq}}$ and $n_{\mathrm{eq}}$ are the equilibrium abundance and number density of $\dm$, respectively. We identify $\Gamma_{A}\equiv n_{\mathrm{eq}}\avg{\sigma\abs v}$ as the annihilation rate of $\dm$ when in equilibrium. The thermally-averaged cross section can be obtained as \citep{Gondolo:1990dk}
\begin{equation}
    \avg{\sigma\abs v} =
        \frac{
            \int_{s_{\mathrm{min}}}^\infty\du s\left(
                s-4m_\dm^2
            \right)\sqrt s\sigma K_1\left(\sqrt s/T\right)
        }{
            8m_\dm^4TK_2(m_\dm/T)^2
        }.
\end{equation}
It is clear from \cref{eq:boltzmann-abundances} that the abundance will stabilize once $\Gamma_A/H\lesssim1$. This condition gives an estimate of the temperature $\TFO$ at which $\dm$ departs from equilibrium, and thus allows us to test whether a set of parameter values is consistent with BBN observables.

In particular, we can immediately estimate the impact of changing the threshold used for assessing BBN constraints. Since the DM is relativistic at decoupling, the freeze-out temperature can be estimated by the relation $T^3\left\langle\sigma\abs v\right\rangle\sim T^2/M_{\mathrm{Pl}}$, where $\sigma$ is the DM annihilation cross section. For our operators, the cross sections scale like $s/\supp^4$ or $1/\supp^2$, so if we adjust $\TFO$ and determine the corresponding value of $\supp$, then $\supp$ is approximately proportional to $\TFO^{3/4}$ or $\TFO^{1/2}$. In particular, we expect the difference between the \BBNThresholdT~threshold and the \SI{2.3}{\MeV} threshold to correspond to a $\mathcal O(1)$ factor in the constraint on $\supp$.

In general, when studying the decoupling of $\dm$, it is important to consider the coupling to neutrinos as well as electrons. If $\dm$ has a non-negligible coupling to neutrinos, it is conceivable that the DM could be kept in equilibrium at later times via thermal contact with the neutrino bath, which would tend to strengthen our constraints. However, the coupling to neutrinos can always be set to zero independent of the coupling to electrons: we assume $\dm$ couples to the neutrino only via the $\operatorname{SU}(2)_L$ doublet, and $\dm$ can couple independently to $e_R$ and to $e_L$. Thus, when evaluating BBN constraints, we ignore thermal contact with neutrinos in order to obtain the most conservative limits.

\subsection{Effective number of neutrinos in CMB}
Another powerful constraint applicable to a new light species is the effective number of neutrino species, $\Neff$, as measured from CMB. To establish constraints with the greatest possible generality, we evaluate bounds from the CMB without regard to the BBN constraints. As we will show, the bounds from BBN and the CMB are comparable in reach, but imposing each independently means that exceptional cases that escape one bound or the other can still be constrained.

$\Neff$ characterizes the contributions to the radiation energy density at recombination from relativistic species apart from photons, and is defined by 
\begin{equation}
    \label{eq:neff}
    \frac{\rho_{\mathrm{rad}}}{\rho_\gamma}\equiv
        1+\frac78\left(\frac{4}{11}\right)^{4/3}\Neff.
\end{equation}
In the absence of any other relativistic species, $\Neff\simeq 3$. The SM actually predicts $\Neff=3.046$, accounting for the three neutrino species and for small effects due to non-idealities in the decoupling process \citep{Mangano:2001iu,Mangano:2005cc}. This is consistent with analyses of Planck data, which find $\Neff\simeq3.1\pm0.2$ \citep{Ade:2015xua}. Additional species are strongly disfavored. A single additional relativistic degree of freedom, i.e., a real scalar, is weakly consistent with current limits. However, CMB stage 4 experiments are expected to measure $\Delta\Neff\equiv\Neff-3.046$ to within $\pm0.03$, which is just sensitive enough to probe the minimum contribution from a real scalar at $1\sigma$ \citep{Abazajian:2016yjj}.

But a new species need not be relativistic at recombination to alter $\Neff$. The introduction of a light DM species can change $\Neff$ by modifying the ratio of the photon and neutrino temperatures \citep{Ho:2012ug,Boehm2013,Brust:2013xpv,Green2017}, and hence the ratio of energy densities in \cref{eq:neff}. In the absence of additional species, the chemical decoupling of electrons and neutrinos takes place at $T_D^0\approx\SI{2.3}{\MeV}$ \citep{Enqvist:1991gx}. Any entropy transferred from DM to electrons after this decoupling leads to heating of the photon bath, and any entropy transferred to neutrinos heats the neutrino bath. If the new species transfers entropy differentially to the photon and neutrino baths at any time after the two baths decouple, the temperature ratio of the baths is modified. Note that $\Delta\Neff$ thus depends on the relative size of the couplings to electrons and neutrinos, as pointed out in \cite{Ho:2012ug} and detailed extensively in \cite{Escudero:2018mvt}.

Typically, the DM will transfer its entropy to one or both baths as a consequence of the conservation of comoving entropy density: when the DM becomes non-relativistic while still in thermal equilibrium, the associated entropy must be transferred to any relativistic species to which it is still coupled. Thus, these species are heated when the DM becomes non-relativistic. Now, suppose that a sub-MeV DM species is coupled to electrons and neutrinos when $T<T_D^0$. If the DM species decouples from one and only one of these two relativistic species before it becomes non-relativistic itself, then the DM will reheat only one of the two baths, changing the temperature ratio. An exception to this rule occurs when the DM \emph{enters} equilibrium with one bath below $T_D^0$, so that the DM accepts entropy of the same order that it loses upon decoupling later on \citep{Berlin:2017ftj}. We will discuss this scenario further in \cref{sec:discussion}.

We now examine the calculation of $\Neff$ in detail. We will write $T_{XY}$ to denote the temperature at which species $X$ and $Y$ lose \emph{direct} thermal contact, i.e., the temperature below which $\Gamma(X\leftrightarrow Y)/H<1$ in our effective theory. The species $X$ and $Y$ might be kept in thermal equilibrium by a third species $Z$ in our framework, i.e., through processes $X\leftrightarrow Z$ and $Z\leftrightarrow Y$ that remain active. We define $T_D$ to be the actual temperature at which electrons and neutrinos drop out of thermal equilibrium with one another once all inter-conversion processes have frozen out, including multi-step processes involving the dark species. Thus, in the standard scenario, $T_D=T_{e\nu}\equiv T_D^0\approx\SI{2.3}{\MeV}$, but the introduction of a new species can keep electrons and neutrinos in thermal equilibrium at lower temperatures.

In particular, suppose that DM decouples from electrons instantaneously at a temperature $T_{\dm e}$, and from neutrinos at a temperature $T_{\dm\nu}$. If $T_D^0<\min\{T_{\dm e},T_{\dm\nu}\}$, then any entropy transferred  to either photons or neutrinos can be shared between the two, so DM reheats these species equally, and the standard calculation is unchanged. However, if $T_{\dm e}<T_D^0<T_{\dm\nu}$, then $\dm$ remains in thermal contact with photons while relativistic, reheating the photon bath but not the neutrino bath. This increases the photon temperature, \emph{reducing} $\Neff$. Similarly, if $T_{\dm\nu}<T_D^0<T_{\dm e}$, then the reverse is true: DM reheats the neutrino bath, and $\Neff$ increases.

\begin{figure*}\centering
    \includegraphics[width=\plotscale\textwidth]{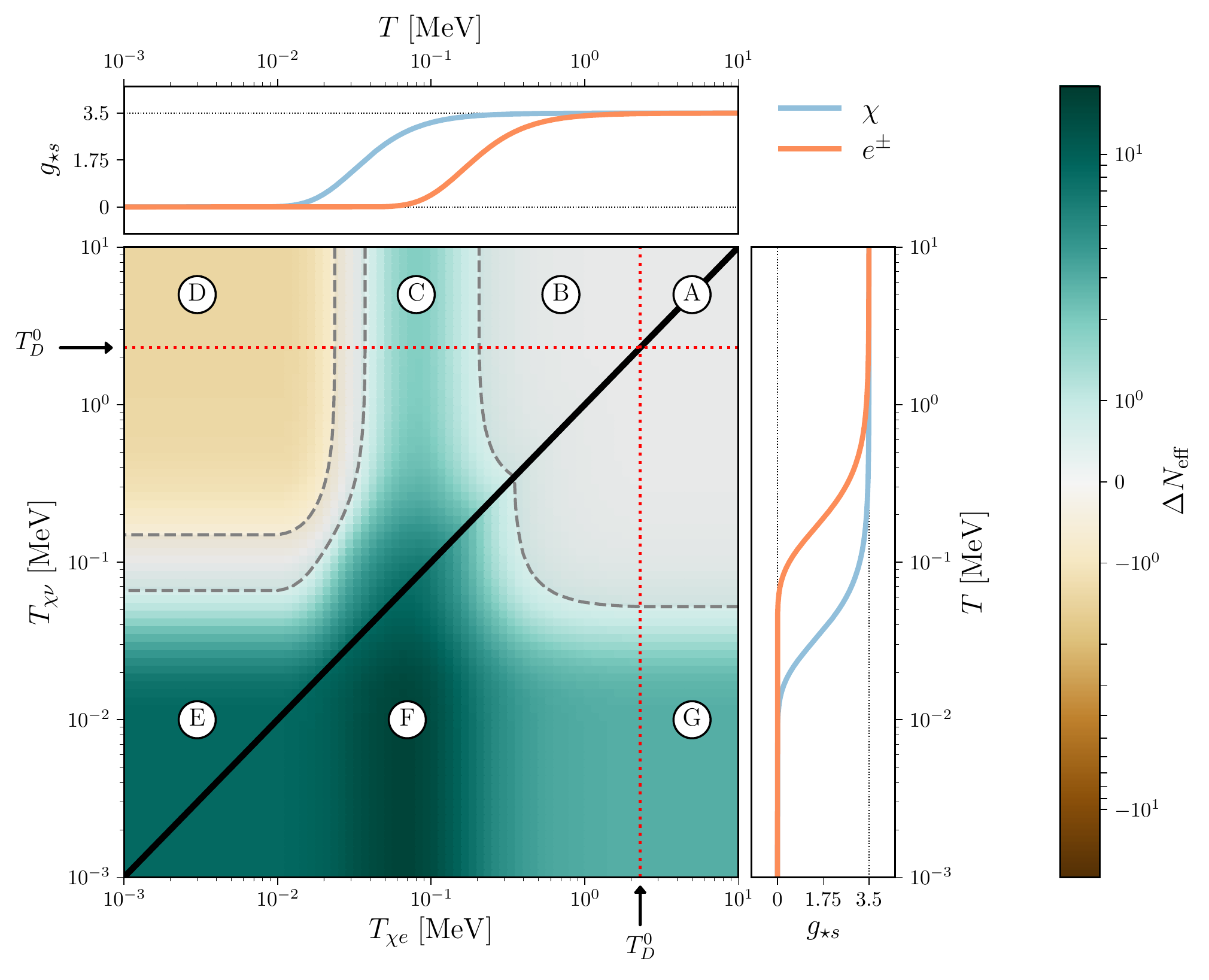}
    \caption{\ignorespaces
        $\Delta N_{\mathrm{eff}}$ as a function of the two decoupling temperatures $T_{\dm e}$ and $T_{\dm\nu}$, assuming that $\dm$ is a Dirac fermion with mass \SI{100}{\kilo\electronvolt}. Side and top panels show entropic degrees of freedom as a function of temperature. Gray shaded area indicates the region consistent with current data at $2\sigma$. Labeled regions can be understood qualitatively as follows.
        \textit{Region A:} $T_{\dm e},T_{\dm\nu}>T_D^0$. Thus any entropy transferred by $\dm$ is shared between the $\gamma$ and $\nu$ baths before they decouple. The standard calculation of $N_{\mathrm{eff}}$ is unaltered.
        \textit{Region B:} $T_{\dm e}<T_D^0<T_{\dm\nu}$. However, $\dm$ and $e^\pm$ are relativistic at both decoupling events, so little entropy is transferred to either the $\gamma$ or the $\nu$ bath.
        \textit{Region C:} Now $e^\pm$ becomes non-relativistic while still in thermal contact with the relativistic $\dm$. The entropy ordinarily transferred by $e^\pm$ to $\gamma$ is now shared with $\dm$, so $\gamma$ is reheated less efficiently, and $\Neff$ increases.
        \textit{Region D:} Here $\dm$ is relativistic below both $T_D^0$ and $T_{\dm\nu}$, but becomes non-relativistic before $T_{\dm e}$ is reached. Thus, $\dm$ reheats the $\gamma$ bath exclusively upon becoming non-relativistic, decreasing $\Neff$.
        \textit{Region E:} $\dm$ becomes non-relativistic above both $T_{\dm\nu}$ and $T_{\dm e}$, so it reheats both baths. The impact on $\Neff$ in this region comes from the delayed $e^\pm$--$\nu$ decoupling (see text).
        \textit{Region F:} $T_{\dm e}>T_{\dm\nu}$, and $\dm$ is relativistic at $T_{\dm e}$. Thus, in addition to the delayed $e^\pm$--$\nu$ decoupling, $\dm$ reheats the $\nu$ bath.
        \textit{Region G:} The electron and $\dm$ are relativistic at $T_{\dm e}$, so here the impact on $\Neff$ is due to $\dm$ reheating the $\nu$ bath.
    }
    \label{fig:neff-combined}
\end{figure*}

The only other possibility is $\max\{T_{\dm e},T_{\dm\nu}\}<T_D^0$, in which case $\dm$ acts as a thermodynamic mediator between electrons and neutrinos below $T_D^0$. In this situation, electrons and neutrinos remain in thermal equilibrium until the temperature falls below $T_D=\max\left\{T_{\dm e},T_{\dm\nu}\right\}$. If the electron is still relativistic throughout this process, then the impact on $\Neff$ is determined by the ordering of $T_{\dm e}$ and $T_{\dm\nu}$. But if $T_D\lesssim m_e$, the impact on $\Neff$ is quite different: photons and neutrinos are still in thermal contact while electrons become non-relativistic, so the \emph{electron} also transfers some of its entropy to the neutrino bath. As we will see shortly, this can have a dramatic impact on $\Neff$.

To calculate $\Neff$, we follow the procedure described in \cite{Boehm:2012gr}. In our scenario, the DM species is non-relativistic at recombination, so we assume that $\Neff$ is not modified by any additional degrees of freedom \emph{at recombination}. Then, given the temperature ratio of the neutrino and photon baths at recombination, $\Neff$ is given by
\begin{equation}
    \Neff = \left(\frac{4}{11}\right)^{-4/3}
        \left(\left.\frac{T_\nu}{T_\gamma}\right|_{\mathrm{rec}}\right)^4N_\nu,
\end{equation}
where $N_\nu$ is the number of SM neutrinos (3). In turn, we can determine the temperature ratio from conservation of comoving entropy density.

Recall that the entropy density of a relativistic bosonic species $i$ with $g_i$ internal degrees of freedom is given by $2\pi^2g_iT^3/45$. Away from the relativistic limit, denoting the true entropy density by $s_i$, we say that this species has $g_{\star s} \equiv s_i/(2\pi^2T^3/45)$ entropic degrees of freedom. Now, let $g_{\star s}^{(\gamma)}$ and $g_{\star s}^{(\nu)}$ denote the entropic degrees of freedom in equilibrium with photons and neutrinos, respectively. Then $g_{\star s}^{(\alpha)}$ is given explicitly by
\begin{equation}
    g_{\star s}^{(\alpha)}
        = \sum_{i\in I}\frac{15g_i}{4\pi^4}
            \int_{x_i}^{\infty}\du u\,
                \frac{\left[4u^2-x_i^2\right]
                \left[u^2-x_i^2\right]^{1/2}}
                {\exp(u)\pm 1},
\end{equation}
where $x_i=m_i/T_\alpha$, and $I$ indexes all species in equilibrium with species $\alpha$ ($\gamma$ or $\nu$). The sign in the denominator is determined by the statistics of species $i$. It can be shown \cite{Boehm:2012gr} that if no entropy leaves the photon or neutrino baths after they decouple, then
\begin{equation}
    \label{eq:temperature-ratio}
    \left.\frac{T_\nu}{T_\gamma}\right|_{\mathrm{rec}} = \left(
        \left.\frac{g_{\star s}^{(\nu)}}{g_{\star s}^{(\gamma)}}
            \right|_{T_D}
        \left.\frac{g_{\star s}^{(\gamma)}}{g_{\star s}^{(\nu)}}
            \right|_{\mathrm{rec}}
    \right)^{1/3}.
\end{equation}
However, in our scenario, it is possible for entropy to leave one of the two baths below $T_D$: suppose the DM decouples from one of the two baths above $T_D$, and decouples from the other below $T_D$, but while still relativistic. At this second decoupling, the DM's remaining entropy leaves the bath to which it was last coupled. This only happens if $T_{\dm e}<T_D\leq T_{\dm\nu}$ or $T_{\dm\nu}<T_D\leq T_{\dm e}$.

To account for this possibility, we modify the calculation of the temperature ratio as follows. Let us assume for the moment that $T_{\dm e}<T_D\leq T_{\dm\nu}$. Conservation of comoving entropy density in a thermal bath $\alpha$ amounts to the assertion that $g_{\star s}^{(\alpha)}|_TT^3a^3$ is constant, where $a$ is the scale factor. For $T<T_D$, comoving entropy density is conserved in each bath except when $T_\gamma = T_{\dm e}$, so the temperatures of the two baths satisfy
\begin{equation}
    \begin{array}{l}
        T_\nu=k_1a^{-1}{g_{\star s}^{(\nu)}}|_{T_\nu}^{-1/3},
        \\[0.2cm]
        T_\gamma = \begin{cases}
            k_2a^{-1}{g_{\star s}^{(\gamma)}}|_{T_\gamma}^{-1/3}
                & T_{\dm e}<T_\gamma<T_D\\
            k_3a^{-1}{g_{\star s}^{(\gamma)}}|_{T_\gamma}^{-1/3}
                & T_\gamma<T_{\dm e},
        \end{cases}
    \end{array}
\end{equation}
where the $k_i$ are constants. Generally, $T_{\mathrm{rec}} < T_{\dm e}$, so
\begin{equation}
    \left.\frac{T_\nu}{T_\gamma}\right|_{\mathrm{rec}}
        = \frac{k_1}{k_3}\left(
            {g_{\star s}^{(\nu)}}\middle/
            g_{\star s}^{(\gamma)}\middle|_{\mathrm{rec}}
        \right)^{-1/3}.
\end{equation}
Thus, to determine the temperature ratio, it is sufficient to identify the ratio $k_1/k_3$, which can be done in two stages. First, since $T_\nu$ and $T_\gamma$ are equal at $T_D$, we must have
\begin{equation}
    \frac{k_1}{k_2} = \left(
            {g_{\star s}^{(\nu)}}\middle/
            g_{\star s}^{(\gamma)}\middle|_{T_D}
        \right)^{1/3}.
\end{equation}
Similarly, at $T_{\dm e}$, $g_{\star s}^{(\gamma)}$ changes discontinuously while $T_\gamma$ is continuous in $a$. Thus, $k_3$ must satisfy
\begin{equation}
    \frac{k_3}{k_2} = \left(
        \frac{
            g_{\star s}^{(\gamma)}\bigr|_{T_{\dm e}^-}
        }{
            g_{\star s}^{(\gamma)}\bigr|_{T_{\dm e}^+}
        }
    \right)^{1/3},
\end{equation}
where $T_{\dm e}^\pm$ denotes a temperature just above or below $T_{\dm e}$. Now we have
\begin{equation}
    \label{eq:temperature-ratio-corrected}
    \left.\frac{T_\nu}{T_\gamma}\right|_{\mathrm{rec}} = \left(
        \left.\frac{g_{\star s}^{(\nu)}}{
            g_{\star s}^{(\gamma)}}\right|_{T_D}
        \frac{
            g_{\star s}^{(\gamma)}\bigr|_{T_{\dm e}^+}
        }{
            g_{\star s}^{(\gamma)}\bigr|_{T_{\dm e}^-}
        }
        \left.
            \frac{g_{\star s}^{(\gamma)}}{g_{\star s}^{(\nu)}}
        \right|_{\mathrm{rec}}
    \right)^{1/3}.
\end{equation}
A similar calculation applies if $T_{\dm\nu}<T_D\leq T_{\dm e}$. Note that \cref{eq:temperature-ratio-corrected} still assumes that $\dm$ does not \emph{enter} equilibrium below $T_D$, an exception we discuss further in \cref{sec:discussion}.

From \cref{eq:temperature-ratio-corrected}, it is easy to see why low DM decoupling temperatures can have a large impact on $\Neff$. In the standard scenario, $g_{\star s}^{(\gamma)}|_{T_D}$ includes photons (2) and relativistic electrons ($\frac78\times 4$), which gives
\begin{equation}
    \frac{g_{\star s}^{(\gamma)}|_{\mathrm{rec}}}
        {g_{\star s}^{(\gamma)}|_{T_D}}
        = \frac{2}{2+\frac78\times4} = \frac4{11}.
\end{equation}
But if neutrinos and photons remain in thermal contact after electrons become non-relativistic, then $g_{\star s}^{(\gamma)}|_{T_D}$ includes only photons, and the above ratio is increased to 1. This increases $\Neff$ by a factor of $(11/4)^{4/3}\approx3.9$, already leading to $\Neff\approx 12$. If $T_{\dm e}<T_{\dm\nu}$, then $\dm$ reheats the photon bath when it becomes non-relativistic, reducing $\Neff$. But if $T_{\dm\nu}<T_{\dm e}$, then $\dm$ reheats the neutrino bath, increasing $\Neff$ even further. The impact of relative decoupling temperatures on $\Neff$ is shown in \cref{fig:neff-combined}.

This approach assumes that the decouplings take place instantaneously, which is generally a good approximation. However, the approximation is poor when the decoupling process overlaps the range of temperatures during which a species becomes non-relativistic. In this case, the entropy of the species is changing rapidly, so it is difficult to estimate the amount of entropy transferred to other relativistic species before decoupling is complete. The temperature ratio can be determined precisely by numerical methods \cite[see e.g.][]{Escudero:2018mvt,Escudero:2020dfa}, and while that lies outside the scope of the present work, we note that instantaneous decoupling should be an effective approximation away from a narrow range of temperatures $T_{\dm e}$ and $T_{\dm\nu}$, corresponding to a very small span of $\supp$ values in our parameter space.

To translate these results into constraints on the coupling between $\dm$ and electrons, we must make an assumption about the coupling between $\dm$ and neutrinos. If the coupling to neutrinos is very small, then $\dm$ may maintain thermal contact with electrons after decoupling from neutrinos. On the other hand, if the coupling to neutrinos is very large, then $\dm$ may remain in thermal contact with neutrinos after decoupling from electrons. In our case, we will assume that $\dm$ couples to $\nu$ exclusively by coupling to the lepton doublet $(e_L,\nu_e)^{\mathrm T}$. That is, we will assume that the $\dm$--$\nu$ coupling is the same as the $\dm$--$e_L$ coupling.

Even in this framework, the impact on $\Neff$ depends on the relative strengths of the $\dm$--$e_L$ and $\dm$--$e_R$ couplings. A non-zero coupling to $e_R$ tends to keep $\dm$ in equilibrium with electrons to lower temperatures, meaning that $\dm$ typically reheats the photon bath. This reduces the temperature ratio of \cref{eq:temperature-ratio}, producing $\Delta\Neff < 0$. However, if $\dm$ stays in equilibrium long enough to modify $T_D$, then we can obtain $\Delta\Neff>0$, as discussed above. Either way, increasing the coupling to $e_R$ only strengthens the effect, so we neglect this coupling to obtain conservative constraints. Note that this is different from our assumption in evaluating BBN constraints, where conservative constraints are obtained by neglecting the coupling to $e_L$.

\subsection{Non-thermal production}

A viable model of DM must (partially) account for, but not exceed, the observed DM density of $\Omega_{\mathrm{DM}}h^2\simeq0.12$ \cite{Aghanim:2018eyx}. If the DM is produced by thermal freeze-out, then a larger annihilation cross section reduces the relic density, so larger couplings conducive to direct detection are less likely to overproduce DM. But in the alternative scenario, if DM is produced out of equilibrium, the relic density increases with the annihilation cross section. In this case, overproduction is an important consideration.

If the DM species never attains thermal equilibrium with the SM, the abundance of DM will evolve toward its equilibrium value, but once $\Gamma_A/H\lesssim1$, the abundance will stay fixed. For renormalizable interactions, this out-of-equilibrium production process is the standard freeze-in mechanism \citep{Hall:2009bx}. Out-of-equilibrium production has also been studied for non-renormalizable operators in the context of so-called ultraviolet freeze-in \citep{Elahi:2014fsa}. For temperatures below $\sim\SI{10}{\MeV}$, within the constraints of our framework, such non-thermal production represents the only alternative to the freeze-out scenario.

The relic density of non-thermal DM is determined using the Boltzmann equation, much like the freeze-out case. The only difference is that the DM species $\dm$ is not in thermal equilibrium with $e^\pm$, and thus we cannot assume that $\dm$ has an equilibrium phase space density. Instead, we assume that the density of $\dm$ is negligible, such that the $f_\dm^2$ term drops out of the Boltzmann equation. In other words, starting from \cref{eq:boltzmann-abundances}, we approximate $Y/Y_{\mathrm{eq}}\simeq0$, which gives $Y^\prime(x) \simeq n_{\mathrm{eq}}(x)\left\langle\sigma|v|\right \rangle(x)/H(x)$. It follows that the out-of-equilibrium yield can be estimated as
\begin{equation}\label{eq:1d-relic}
    Y(\infty) \simeq
        Y(x_{\mathrm{min}}) + \int_{x_{\mathrm{min}}}^\infty\du x\,\frac{
            n_{\mathrm{eq}}(x)
            \left\langle\sigma|v|\right\rangle(x)
        }{H(x)}.
\end{equation}
As with freeze-out, the relic density in the non-thermal case is determined by the DM mass and couplings with SM particles. However, there is also a dependence on initial conditions in the form of $x_{\mathrm{min}}$ and $Y(x_{\mathrm{min}})$. In the freeze-out scenario, the abundance of DM in the early universe is simply the equilibrium abundance: equilibrium effectively erases the initial condition. But in the non-thermal scenario, equilibrium is never attained, so the dependence on the initial abundance is retained. Typically, when DM is produced by SM annihilations out of equilibrium, one calculates the relic density by fixing the DM density to zero at very early times and evolving non-thermally. This procedure requires that the interactions considered are renormalizable, in order for the production process to be modeled consistently at very high temperatures. Our effective operators are non-renormalizable, so we cannot determine the relic density precisely in the non-thermal case: the result depends on the choice of UV completion.

However, we can still place a lower bound on the relic density. We require that our effective theory is valid at scales below $\sim\SI{10}{\MeV}$, so if we fix the abundance to some value at $\SI{10}{\MeV}$, we can determine the resulting relic abundance. In particular, by fixing the initial abundance to zero, we necessarily underestimate the relic density. This corresponds to a choice of $x_{\mathrm{min}}$ and the condition that $Y(x_{\mathrm{min}})=0$. With this initial condition, we can exclude models on the basis of their relic densities even when they never attain thermal equilibrium with the SM. Further, these constraints are determined entirely by conditions below $\TBBN$, and are thus completely independent of the UV completion.

Note that if $\supp$ is sufficiently small, then even with this initial condition, the DM species will thermalize with the SM between $\TBBN$ and the present day. In this case, the relic density is set by the standard freeze-out paradigm, and \cref{eq:1d-relic} is not valid. Even if the DM species does not quite enter thermal equilibrium, as long as it attains a non-negligible abundance, \cref{eq:1d-relic} can significantly overpredict the relic density. Thus, while \cref{eq:1d-relic} is useful to understand the qualitative features of the non-thermal relic density, we evaluate the constraint by numerically solving \cref{eq:boltzmann-abundances}.

As in the previous cases, we need to specify the coupling to neutrinos to perform these calculations consistently. Since the neutrino bath has a temperature comparable to the electron bath, a light $\dm$ can be effectively produced by neutrinos as well as electrons. Thus, a coupling between $\nu$ and $\dm$ can significantly affect the relic abundance. However, as with the coupling to electrons, the relic density is not monotonic in the coupling to neutrinos. If the DM never enters thermal equilibrium with any SM species, then a coupling to neutrinos tends to enhance the relic abundance by providing another production channel. On the other hand, if DM does enter equilibrium \emph{with neutrinos}, then a larger coupling to neutrinos keeps it in equilibrium longer, reducing the relic abundance. However, at most of the points of interest in our parameter space, the constraint is driven by out-of-equilibrium production, so we neglect the coupling to neutrinos when evaluating the relic density.

\section{Constraints and detection rates}
\label{sec:constraints}
The constraints we place on sub-MeV DM are relevant for direct detection experiments based on elastic electron--DM recoils. In principle, there are many such experiments, but they share several important features. Generically, electron recoil experiments prepare a low-temperature collection of electrons for scattering with galactic halo DM, and by whatever mechanism, the experiment is sensitive to deposited recoil energies between some $E_{\mathrm{min}}$ and $E_{\mathrm{max}}$. We calculate the detector sensitivity following \cite{Hochberg2016a}, but the results are typical of electron recoil experiments with very low thresholds.

\subsection{Estimation of the event rate}
In the proposal of \cite{Hochberg2016a}, the detector is constructed from an aluminum superconductor. At low temperatures, electrons move through the detector with velocities of order the Fermi velocity $v_F$, and with the appropriate instrumentation, recoil energies as low as \SI{1}{\milli\electronvolt} may be detectable. We now review the calculation of the detection rate, following \cite{Hochberg2016a} and \cite{Reddy1997}.

To compute the detection rate, we will consider scattering events at fixed recoil energy $E_R$. We label the initial and final DM momenta by $\bb p_1$ and $\bb p_3$, and the initial and final electron momenta by $\bb p_2$ and $\bb p_4$. We do the same for the energies, so that $E_R = E_1 - E_3 = E_4 - E_2$. We define the 3-momentum transfer by $\bb q=\bb p_1 - \bb p_3$. We denote 4-momenta by $P_i$, and we write $q=\abs{\bb q}$ and $p_i=\abs{\bb p_i}$. We denote the local DM number density by $n_\dm$, and the scattering rate by $\Gamma=\left\langle n_e\sigma v_{\mathrm{rel}}\right\rangle$. The event rate per unit detector mass is
\begin{equation}
    \label{eq:rate}
    R = \frac{n_\dm}{\rho_{\mathrm{detector}}}
        \int\du v_\dm\dd E_R\,f_\dm(v_{\dm})\,
            \frac{\du\Gamma(v_\dm,E_R)}{\du E_R},
\end{equation}
where $f_\dm(v_\dm)$ is the local DM velocity distribution in the lab frame. We take the velocity distribution to be a Maxwell--Boltzmann distribution in the galactic frame with rms velocity \SI{220}{\kilo\meter/\second} and a cutoff at the halo escape velocity $v_{\mathrm{esc}}\simeq\SI{500}{\kilo\meter/\second}$. We then determine $f_\dm(v_\dm)$ by taking the Earth velocity to be \SI{244}{\kilo\meter/\second} in the galactic frame \cite{Lewin:1995rx}.

Now we turn to the evaluation of the scattering rate $\Gamma(v_\dm,E_R)$. Observe that $\Gamma$ not only contains the scattering cross section, but also accounts for the effects of Pauli blocking, effectively controlling the available phase space for scattering events. Following \cite{Reddy1997}, we estimate $\Gamma$ by
\begin{multline}
    \label{eq:scattering-rate}
    \frac{\du\Gamma(E_1,E_R)}{\du E_R} = \int
        \frac{\du^3\bb p_2}{(2\pi)^3}
        \frac{\du^3\bb p_3}{(2\pi)^3}
        \frac{\du^3\bb p_4}{(2\pi)^3}
            \, W(\bb p_1,\bb p_2,\bb p_3,\bb p_4) \times \\
        2f_{\mathrm{FD}}(E_2)
        (1-f_{\mathrm{FD}}(E_4))\delta_E\delta_P^4,
\end{multline}
Here, $\delta_P^4$ is a Dirac delta enforcing conservation of 4-momentum; $\delta_E$ fixes the recoil energy, setting $E_1-E_3=E_R$; $f_{\mathrm{FD}}(E) = 1/(1+\exp(E-\mu)/T)$ is the Fermi-Dirac distribution; and we define
\begin{equation}\label{eq:W-definition}
    W(\bb p_1,\bb p_2,\bb p_3,\bb p_4) = \frac{
        \bigl\langle\abs{\mathcal M}^2\bigr\rangle
    }{16E_1E_2E_3E_4},
\end{equation}
where $\bigl\langle\abs{\mathcal M}^2\bigr\rangle$ is the matrix element for the scattering process.

In many cases of interest, $W$ is independent of the initial and final momenta of the target ($\bb p_2$ and $\bb p_4$), in which case the rate factorizes as 
\begin{equation}\label{eq:rate-factorization}
    \frac{\du\Gamma(E_1,E_R)}{\du E_R} = \int\frac{\du^3\bb p_3}{(2\pi)^3}
        \delta_EW(\bb p_1,\bb p_3)S(E_R,q),
\end{equation}
where  $S$ accounts for Pauli blocking, and is given explicitly by
\begin{equation}
    S(E_R,q) = \int\frac{2\dd^3\bb p_2\dd^3\bb p_4}{(2\pi)^2}
        f_{\mathrm{FD}}(E_2)(1-f_{\mathrm{FD}}(E_4))\delta_P^4.
\end{equation}
In our EFT, $W$ is not generally independent of the target momenta. However, we can treat scattering in the non-relativistic limit, where such independence is guaranteed: the denominator in \cref{eq:W-definition} is independent of the momenta to first order, and can be replaced with $16m_\dm^2m_e^2$. The squared matrix element depends on the momenta only through the Mandelstam variables $s$ and $t$, which have non-relativistic limits
\begin{equation}
    s \simeq (m_e + m_\dm)^2,
    \qquad
    t \simeq 2 \bb p_1\cdot\bb p_3,
\end{equation}
so $\bigl\langle\abs{\mathcal{M}}^2\bigr\rangle$ is also independent of $\bb p_2$ and $\bb p_4$ to first order. Thus, for the remainder of this work, we will consider $W$ to be a function of $\bb p_1$ and $\bb p_3$ only, and factorize the rate as in \cref{eq:rate-factorization}.

We work in the low-temperature limit, where $f_{\mathrm{FD}}$ reduces to a Heaviside step function, $f_{\mathrm{FD}}(E_i)=\Theta(E_F-E_i)$, where $E_F\approx\SI{11.7}{\eV}$ is the Fermi energy of aluminum. In this case, $S(E_R,q)$ can be evaluated explicitly. We perform the $\bb p_4$ integral using the 3-momentum--conservation delta function, and we use the remaining energy-conservation delta function to integrate over $\cos\theta_2$. This leaves a 1-dimensional integral,
\begin{multline}
    S(E_R,q) = \frac{m_e}{\pi q}\int p_2\dd p_2\,
        \Theta\left(
            1-\left|
                \frac{2m_eE_R-q^2}{2p_2q}
            \right|
        \right)\times\\
        \Theta\left(E_F-E_2\right)\left[1-\Theta\left(E_F-E_2-E_R\right)\right]
        .
\end{multline}
This integral can be evaluated directly by comparing the arguments of the Heaviside functions. The result is
\begin{equation}
    S(E_R,q) = \frac{m_e\left(
        m_eE_R - E_S^2
    \right)}{\pi q}\Theta\left(
        2m_eE_R - E_M^2
    \right),
\end{equation}
where $E_M^2=\left(2m_eE_R-q^2\right)^2/(4q^2)$ and $E_S^2$ is given by
\begin{equation}
    E_S^2 = \max\left(
            2m_e(E_F-E_R),\,E_M^2
        \right).
\end{equation}

To actually evaluate the rate in \cref{eq:scattering-rate}, we change coordinates to $(E_R,q)$. Since there is no dependence on the azimuthal angle, we obtain
\begin{equation}
    \du^3\bb p_3 = \frac{2\pi m_\dm q}{p_1}\dd q\dd E_R,
\end{equation}
and the limits of integration are $q_- < q < q_+$, where
\begin{equation}
    q_\pm = \sqrt{
        p_1^2 + p_3^2 \pm 2p_1p_3
    }.
\end{equation}
Under this change of coordinates, in the non-relativistic limit, $t\simeq 2p_1^2-2m_\dm E_R-q^2$. In particular, this means that $W$ depends on $\bb p_1$ and $\bb p_3$ only through $q$, $p_1$, and $E_R$. Then the differential scattering rate $\du\Gamma/\du E_R$ in \cref{eq:rate} is given by
\begin{equation}
    \frac{\du\Gamma}{\du E_R} = \frac{m_\dm}{(2\pi)^2p_1}
        \int_{q_-}^{q_+}q\dd q\,W(p_1,E_R,q) S(E_R,q).
\end{equation}

The limits of the $E_R$ integral in \cref{eq:rate} are set by the lower and upper thresholds of the detector, which we take to be \SI{1}{\meV} and \SI{1}{\eV}, respectively. Note that there are kinematical constraints on the minimum DM velocity ($E_1$) required to deliver a given recoil energy $E_R$. Thus, the cutoff in the velocity distribution effectively imposes a maximum $E_R$ at fixed $m_\dm$.

\subsection{Detection prospects and constraints by operator}
\begin{figure}
    \includegraphics[width=\linewidth]{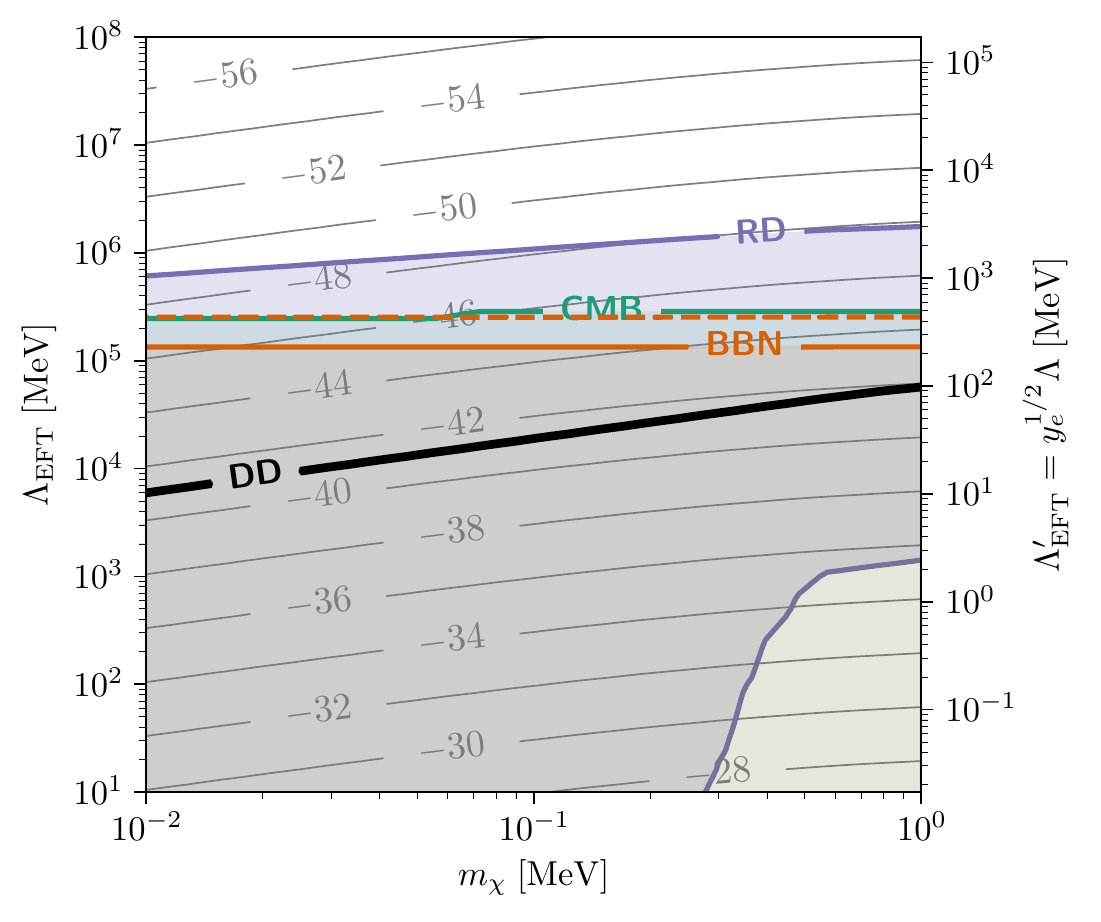}
    \caption{Constraints on a Dirac fermion $\dmf$ interacting via the operator $\fermionoperator{SS}=\supp^{-2}\bar\dmf\dmf\bar ee$ ($\coup=1$).
        Background contours show scattering cross section, labeled as $\log_{10}(\sigma_{\mathrm{scat}}/\SI{}{\centi\meter^2})$.
        \textit{Black, DD:} direct detection sensitivity (95\% CL) with \SI{1}{\kilo\gram.\year} exposure.
        \textit{Green, CMB:} constraint from $\Neff$.
        \textit{Orange, BBN:} solid line: constraint from light element abundances with a threshold temperature of \BBNThresholdT. Dashed line: constraint with a threshold temperature of \SI{2.3}{\MeV} (see \cref{sec:bbn}).
        \textit{Blue, RD:} constraint from relic density.
    }
    \label{fig:single-operator}
\end{figure}

\begin{figure*}[htbp]
    \centering
    \includegraphics[width=\plotscale\textwidth]{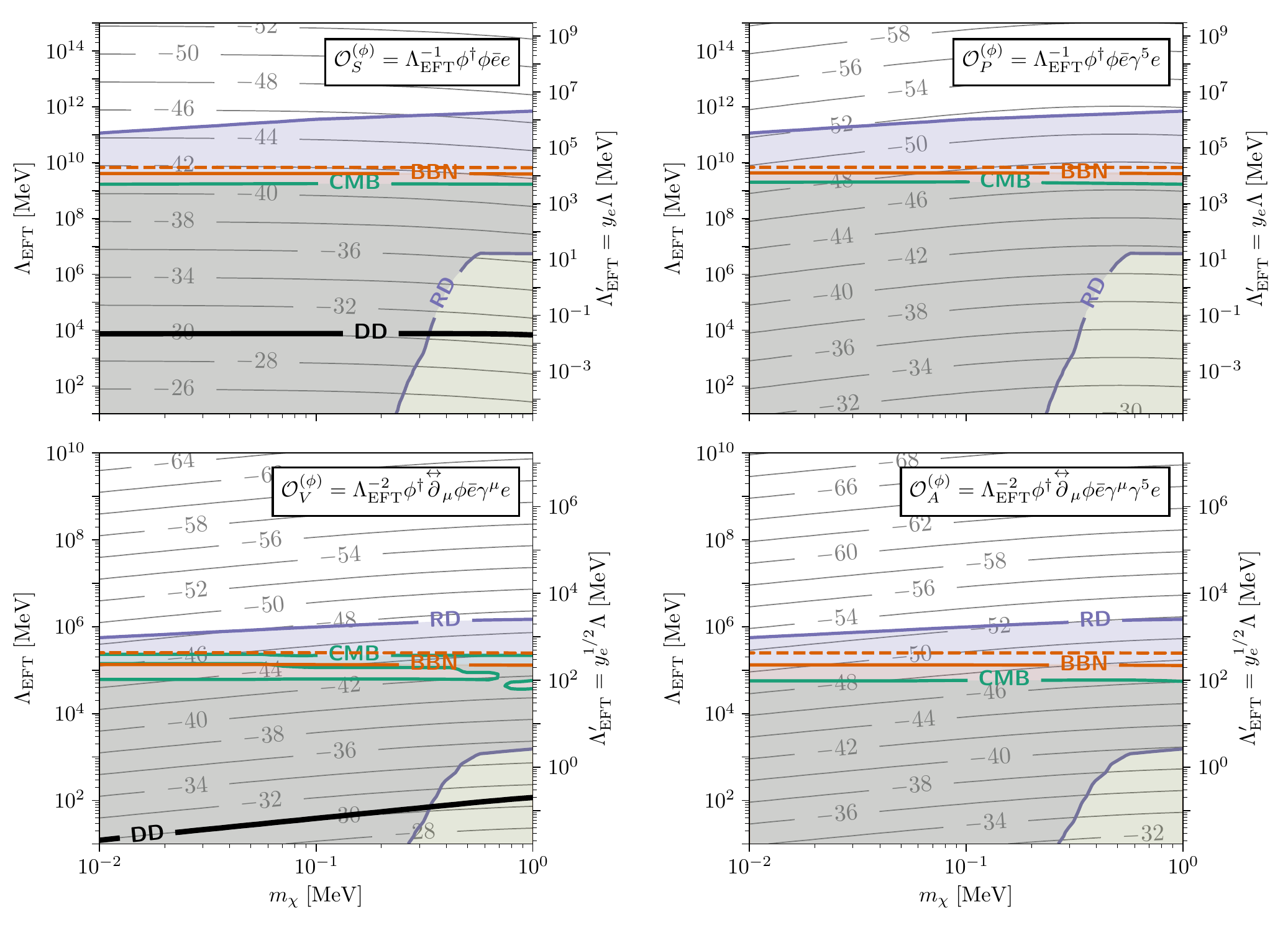}
    \caption{Constraints by operator for DM a scalar $\dms$.
        \constraintlabels
        Note that the direct detection contour does not appear for $\scalaroperator{P}$ or $\scalaroperator{A}$. For these operators, direct detection can constrain smaller values of $\supp$ than shown on the plot, but our framework requires that $\supp\gtrsim\SI{10}{\MeV}$.
        }
    \label{fig:scalar-grid}
\end{figure*}

\begin{figure*}[htbp]
    \centering
    \includegraphics[width=\plotscale\textwidth]{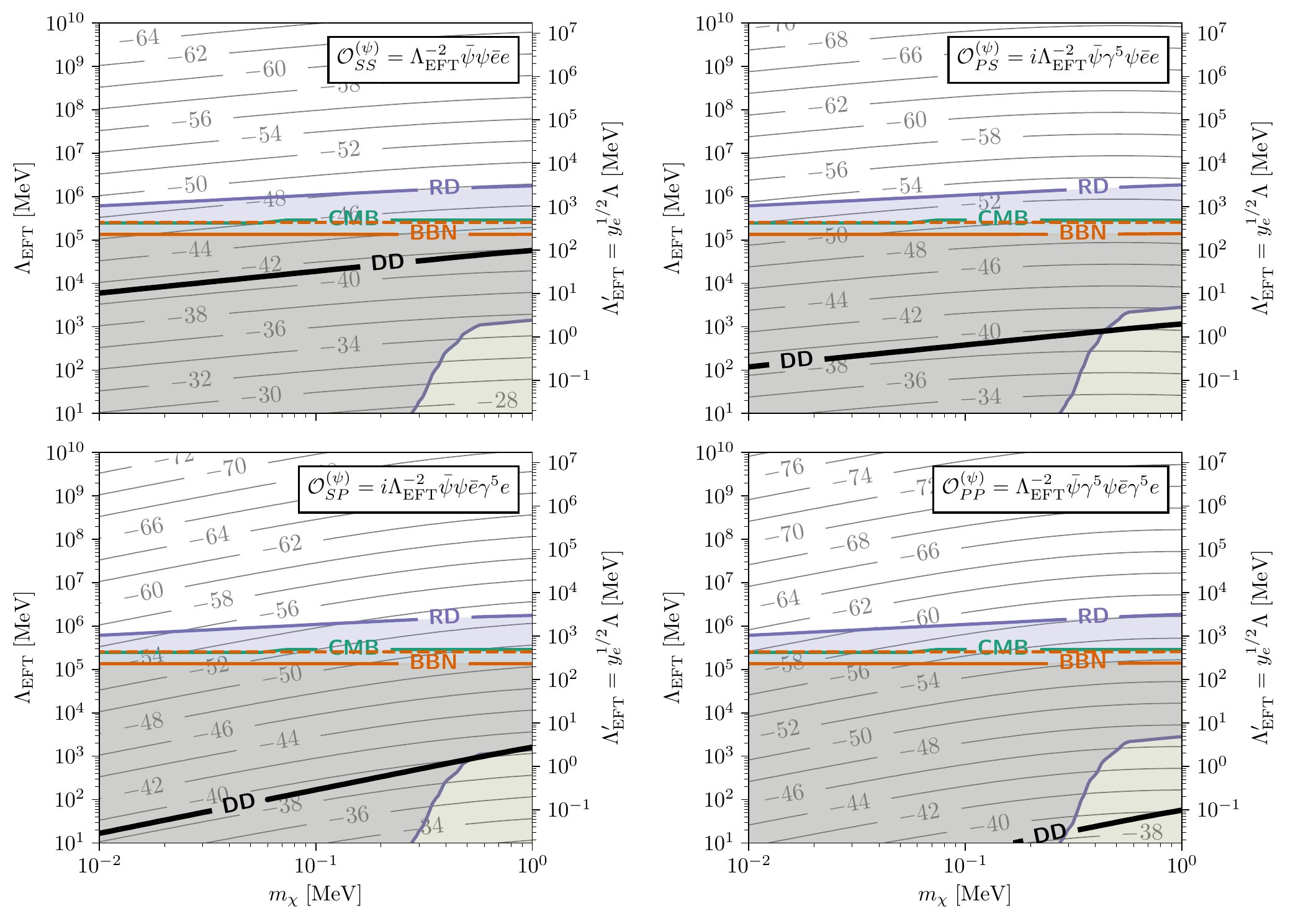}
    \caption{
        Constraints by operator for DM a fermion $\dmf$, for operators composed of scalar or pseudoscalar bilinears.
        \constraintlabels
    }
    \label{fig:fermion-grid-S}
\end{figure*}

\begin{figure*}[htbp]
    \centering
    \includegraphics[width=\plotscale\textwidth]{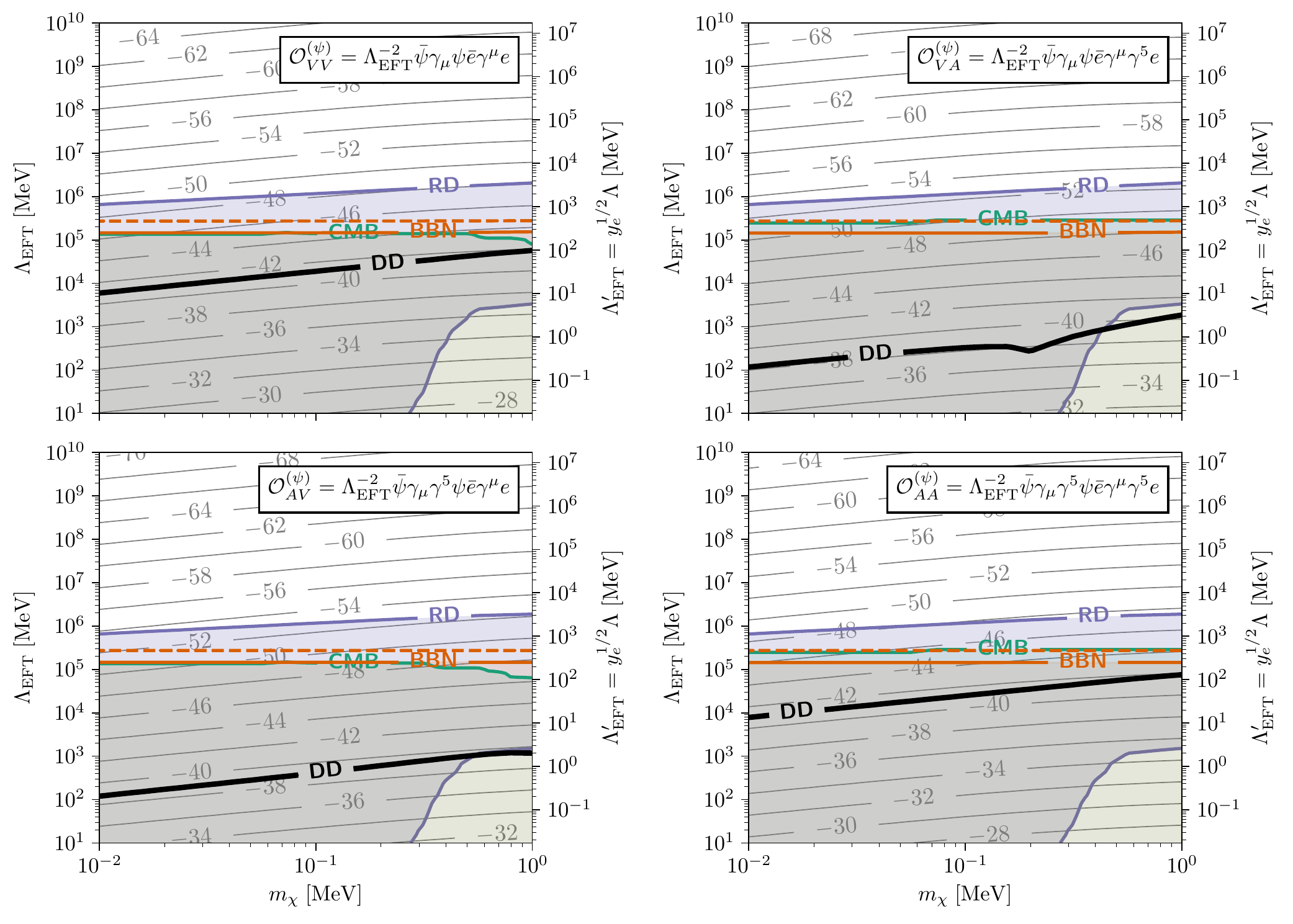}
    \caption{
        Constraints by operator for DM a fermion $\dmf$, for operators containing a vector or axial vector current.
        \constraintlabels
    }
    \label{fig:fermion-grid-V}
\end{figure*}

\begin{figure*}[htbp]
    \centering
    \includegraphics[width=\plotscale\textwidth]{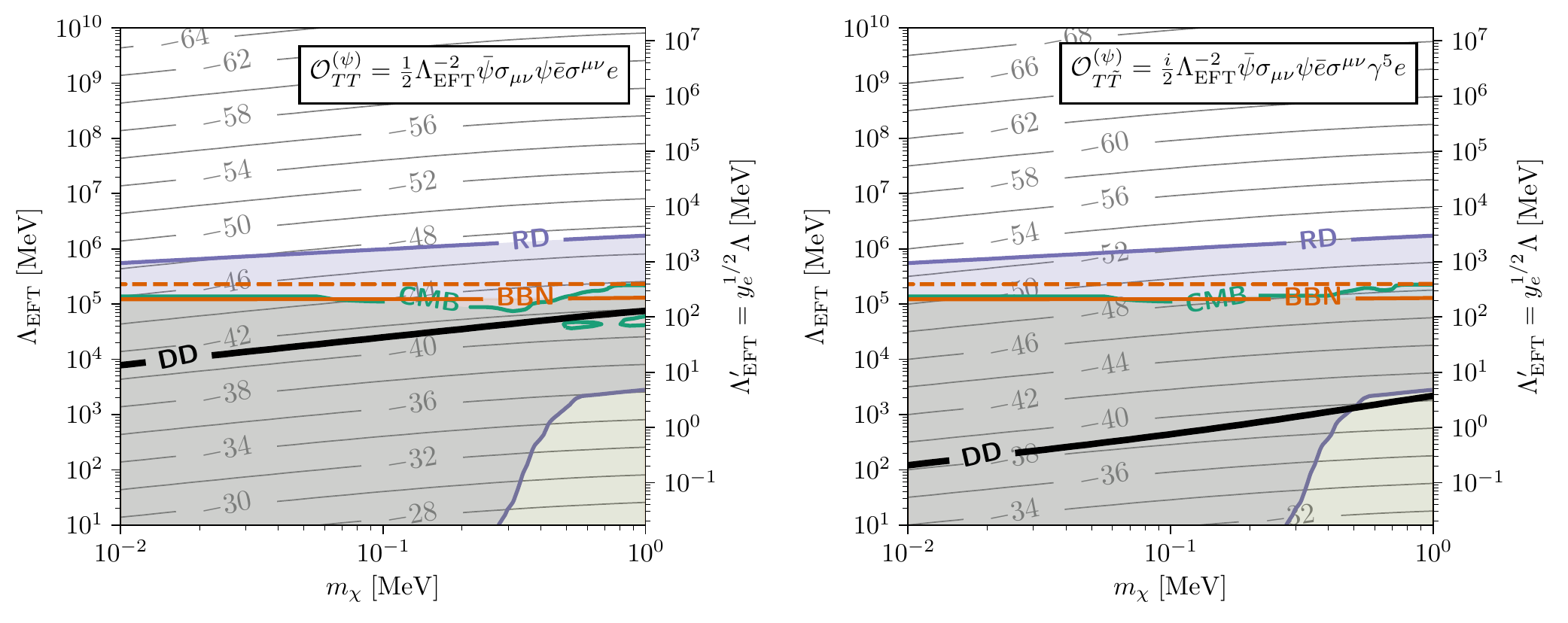}
    \caption{
        Constraints by operator for DM a fermion $\dmf$, for operators containing a spin-2 current.
        \constraintlabels
    }
    \label{fig:fermion-grid-T}
\end{figure*}

We now examine our cosmological constraints in relation to the projected experimental reach for each of the operators in \cref{tab:eft-scalar,tab:eft-fermion}. \Cref{fig:scalar-grid,fig:fermion-grid-S,fig:fermion-grid-V,fig:fermion-grid-T} show cosmological constraints alongside projected 95\% CL direct detection constraints with a \SI{1}{\kilo\gram\year} exposure. In order to point to some general features of our results, we duplicate constraints for $\fermionoperator{SS}$ in \cref{fig:single-operator}. However, the following discussion applies to all of the results in \cref{fig:scalar-grid,fig:fermion-grid-S,fig:fermion-grid-V,fig:fermion-grid-T}.

All of the interactions considered for $\dmf$ a Dirac fermion can also be evaluated for $\dmf$ a Majorana fermion, and we do not consider matrix elements for Majorana fermions separately. Rather, we can directly relate our cosmological constraints on a Dirac fermion to the Majorana case. Whereas the relic density is controlled by $n_\dmf\Gamma_A=n_\dmf^2\avg{\sigma\abs v}$ for a Dirac fermion, this expression double-counts the phase space for a Majorana fermion. Since the relic density is inversely proportional to the annihilation rate, it follows that the relic density of a Majorana fermion is simply twice that of a Dirac fermion with the same mass and interactions \citep{Gondolo:1990dk,Dreiner:2008tw}.

The annihilation rate also sets the freeze-out temperature for a species in equilibrium with the SM, via the condition $\Gamma_A\simeq H$. In general, $\Gamma_A\sim\supp^{4-k}$ for a dimension-$k$ operator. All of our operators with DM a fermion are dimension-6, so to go from the Dirac case to the Majorana case, it is sufficient to make the replacement
$\supp\to 2^{-1/(4-k)}\supp=\sqrt2\supp$. In principle, the value of $\Neff$ is also different in the Majorana case, but in nearly the entire excluded parameter space, $\Delta\Neff$ is large compared with experimental uncertainty, sufficient to rule out a Majorana fermion as well as a Dirac fermion. Thus, in sum, the cosmological constraint curves in \cref{fig:single-operator} are shifted up slightly by a factor of $\sqrt2$ in the Majorana case, while the direct detection projections are unchanged.

In each figure, the left vertical axis shows the suppression scale $\supp$, effectively corresponding to inverse coupling. Thus, a stronger constraint line appears higher on the plot, and excludes the parameter space below. The left axis in each plot gives the value of $\supp$ alone, and the coupling $\coup$ is taken to be 1. This is distinct from fixing $\coup/\supp$ or $\coup/\supp^2$, since we must have $\supp\gg\TBBN$ at all points regardless of the value of the coupling. Otherwise, the EFT would be applied outside its regime of validity.

However, as discusssed in \cref{sec:EFT}, many UV completions naturally generate a coupling of order $y_e$. To account for this possibility, we show a second vertical axis on the right of each plot, corresponding to the value of $\supp$ in the case that $\coup=y_e$. For dimension-5 operators, which appear with a factor of $\supp^{-1}$, this corresponds to $\supp^\prime=y_e\supp$. For dimension-6 operators, $\supp^\prime=y_e^{1/2}\supp$ instead.

Where $\supp^\prime\lesssim\TBBN$, the EFT may not be applicable. This is important, e.g., for comparing the EFT to specific UV completions, but it has little effect on our conclusions: in every case, our constraints become relevant at $\supp^\prime\gg\TBBN$, and a significant range of direct detection cross sections can still be ruled out by cosmology. In principle, cosmological constraints on cross sections that lie below $\supp^\prime\sim\TBBN$ can be evaded by models that have new MeV-scale degrees of freedom in addition to the DM species. However, models of this kind do not generically alleviate the constraints.

The projected direct detection reach (DD, black) is generally the lowest line in each figure, i.e., the weakest constraint. The next line, stronger at low masses by greater than an order of magnitude in $\supp$, is the constraint from light element ratios (BBN, orange). In certain cases, a higher threshold temperature of \SI{2.3}{\MeV} is appropriate, see for instance \cite{Boehm2013} (see \cref{sec:bbn}). The corresponding constraints are shown as dashed curves. However, in general, we can only place a constraint at the lower temperature of \BBNThresholdT, shown with solid curves. In either case, a comparable constraint is obtained from $\Neff$ as measured from $T_\nu/T_\gamma$ (CMB, green). The final constraint is from overproduction of DM (RD, blue). The final constraint is from overproduction of DM (RD, blue). Note that for some operators, there are narrow islands of parameter space where the $\Neff$ constraint is weakened. In these regions, the impact on $\Neff$ is transitioning between $\Delta\Neff<0$ and $\Delta\Neff>0$, as in \cref{fig:neff-combined}. Similarly, some regions with small $\supp$ are not ruled out by overproduction, since the DM thermalizes and freezes out at a lower abundance.

As anticipated in \cref{sec:EFT}, when comparing direct detection prospects to cosmological constraints, no operator improves on the prospects of $\scalaroperator{S}$ for scalar DM. For fermionic DM, on the other hand, we expect that the operators $\fermionoperator{VV}$, $\fermionoperator{AA}$, and $\fermionoperator{TT}$ will be at least competitive with $\fermionoperator{SS}$, and this is borne out by our results. Still, we find no region of parameter space in which the projected direct detection constraints exceed all three cosmological probes for any of our effective operators.

Simplistically, this suggests that any model with a heavy mediator detectable by such an experiment is ruled out by cosmology. However, there remain possible exceptions to these constraints, as we discuss in the following section.

\section{Discussion and conclusions}
\label{sec:discussion}
In this work, we have derived cosmological constraints on a broad class of sub-MeV DM models that can be compared directly with detection prospects in electron recoil detectors. We now revisit the generality of our constraints, point out possible exceptions, and discuss the outlook for sub-MeV DM at electron recoil experiments.

Effectively, our goal has been to derive cosmological constraints on the \emph{scattering} cross section between electrons and sub-MeV DM. Cosmology is mainly sensitive to the DM annihilation cross section, and in order to connect the two cross sections, we have produced these constraints in the context of an EFT. We have enumerated the possible thermal histories for a single DM species in this framework. If the DM is in thermal equilibrium with electrons at high temperatures, then light element abundances and $\Neff$ constrain the freeze-out temperature, and thereby constrain the interactions between $\dm$ and the SM. In the alternative scenario, if the DM is out of equilibrium at early times, a lower bound can be placed on the relic density, providing an independent constraint on the interactions. In both cases, a constraint is placed on the coupling between DM and electrons, assuming a specific form for the interaction.

In general, the form of the operator coupling electrons to DM affects the relationship between the annihilation cross section at early times and the scattering cross section today. Typically, then, constraints obtained by these methods are model-dependent. However, if the DM--SM mediator has a mass above $\sim\SI{10}{\MeV}$, then our approach is quite general: our results are only sensitive to physical processes at lower temperatures, where the EFT is valid and cosmological history is well-established. Still, beyond the mediator mass, there are a few possible exceptions to the constraints derived here.

First, some of these constraints can be evaded with an extended dark sector. In principle, the overproduction constraint can be weakened: such models provide mechanisms to deplete the DM relic density, although we will discuss caveats to this scenario shortly. However, even in this case, the existence of a light DM species is enough for the BBN and $\Neff$ bounds to remain effective---adding additional dark degrees of freedom does nothing to improve the situation. One could still escape these constraints by assuming that a phase transition takes place in an extended dark sector between $\TBBN$ and the present day, such that the EFT is not valid in both epochs.

Another class of exceptions consists of models in which the dark species \emph{enters} thermal equilibrium with the SM below $\TBBN$, and thus below $T_D$, the temperature of neutrino-photon decoupling. In this case, the entropy transferred to the SM bath upon freeze-out can be comparable to the entropy accepted upon equilibration, so the constraint from $\Neff$ can be circumvented \citep{Berlin:2017ftj}. This scenario is possible only in a very limited segment of the heavy-mediator parameter space, which we estimate as follows. We set the abundance of DM to zero at \BBNThresholdT, and then determine the minimum value of $\supp$ below which DM thermalizes before the temperature drops to \SI{0.5}{\MeV}, thus still influencing BBN. Above this value of $\supp$, it is possible to evade bounds from BBN and $\Neff$, depending on initial conditions. Typically, this minimal value of $\supp$ is about one decade weaker than the BBN limit, and still out of reach of direct detection projections across most of our mass range.

Note that the overproduction bound already assumes an initial condition with zero DM abundance, so it cannot be evaded in this way. This is an example of the utility of the several overlapping constraints: the most conservative assumptions are different for each constraint, and correspondingly, exceptions apply differently as well. It is thus necessary to consider all of our constraints simultaneously, even in cases where one constraint appears to dominate. Our goal is to generalize the constraints to the broadest possible class of models, and even though many regions of parameter space are ruled out by multiple observables, it is important to carefully evaluate each constraint independently.

Still, the fact that the overproduction constraint exceeds the constraints from BBN and the CMB is itself a notable result. In general, there are many mechanisms that can influence the dark matter density, so constraints from the relic density are typically confounded by significant model dependence. However, in the scenario of interest, the model dependence is quite limited. To evade the constraint, one would need a mechanism of depleting the dark matter density at temperatures well below \SI{1}{\mega\electronvolt}.

There are some simple methods of accomplishing this depletion, e.g., entropy dilution \cite{Evans:2019jcs}, a late phase transition in the dark sector, or late-time decay of a heavy species into sub-MeV DM today. However, each of these can also be used to evade constraints from BBN and the CMB, so they do not bestow any additional model-dependence on the overproduction bound. It is conceivable that number-changing interactions in the dark sector (e.g. $4\to2$ processes) could be used to deplete the DM density without modifying the other constraints, and this model dependence is unique to the overproduction bound. But even this strategy would only work in a narrow region of parameter space, and in that sense, it is comparable to known exceptions in the usual BBN and CMB bounds \cite{Berlin:2017ftj,Berlin:2019pbq}.

The overproduction constraint thus sets a new target for future direct detection proposals. Considering only BBN and CMB constraints motivates direct detection experiments that probe scattering cross sections a few orders of magnitude beyond the projections in this work. However, overcoming the overproduction bound requires experimental proposals to reach several orders of magnitude beyond the BBN and CMB constraints.

Finally, we note that it might be possible to evade our constraints by taking some arbitrary linear combination of the effective operators in \cref{tab:eft-scalar,tab:eft-fermion}. In principle, in this high-dimensional parameter space, there might be points for which interference of the matrix elements in \cref{tab:scalar-squared-matrix-elements,tab:dirac-squared-matrix-elements} conspires to reduce the DM annihilation or production cross section while preserving the scattering cross section. Then each of our cosmological constraints would be weakened, while the projected direct detection constraints would be maintained. However, in order for this to work, the Wilson coefficients would have to be engineered to produce such a cancellation.

In light of these constraints, the outlook for extant electron recoil detection proposals is brightest for DM masses $\SI{1}{\MeV}\lesssim m_\dm\lesssim\SI{1}{\giga\electronvolt}$ or for mediator masses $m_\dmmed\ll\SI{10}{\MeV}$. In order to access parameter space which is viable in our framework, and in particular to surpass the overproduction bound, future proposals must probe scattering cross sections at least six orders of magnitude beyond current proposals. A light mediator certainly remains a possibility, but is subject to additional constraints \citep[see e.g.][]{Sabti:2019mhn}. The case of a light mediator is thus best studied in the context of simplified models, as in the analysis of \cite{Knapen:2017xzo}. Inelastic scattering may also improve direct detection prospects relative to cosmological constraints, and, of course, DM masses above $\sim\SI{1}{\MeV}$ remain an interesting target. However, if DM is dominantly composed of a single light species, and interacts dominantly with electrons via a heavy mediator, then  cosmological constraints compromise the prospects of proposed experiments.

\acknowledgments

BVL and SP are partially supported by the U.S. Department of Energy grant number DE-SC0010107. This work made use of the \texttt{FeynCalc} package \cite{Shtabovenko:2016sxi,Mertig:1990an} for evaluation of cross sections and the \texttt{ColorBrewer} tool \footnote{\href{http://colorbrewer2.org}{\texttt{http://colorbrewer2.org}}} for selection of colorblind-friendly palettes. We thank Hiren Patel for cross-checking matrix elements and cross sections, and we thank Yonit Hochberg for valuable input regarding superconducting detectors. We thank Miguel Escudero, Rouven Essig, Juri Fiaschi, and Robert Scherrer for comments on an earlier version of this manuscript. We thank the Galileo Galilei Institute for hospitality while parts of this work were completed. We are especially grateful to Francesco D'Eramo, who originally conceived of this project and was deeply involved in most of the execution. Although circumstances compelled Dr. D'Eramo to leave the project later on, we remain deeply appreciative of his input and encouragement. Finally, we are grateful to the anonymous referee for careful feedback and insightful suggestions on earlier versions of this manuscript.

\bibliography{main}

\clearpage

\makeatletter
\let\ftype@table\ftype@figure
\makeatother

\begin{table*}[htbp]
    {
    \setlength\Cwidth{7cm}
    \def\arraystretch{1.7}
    \begin{equation*}
    \begin{array}{|c|@{\hspace{1em}}l@{\hspace{1em}}|}
    \hline
        \text{Operator} &
        \coup^{-2}\supp^{2}\sum_{\text{spin}}
            \left|\mathcal M\right|^2_{\dms\bar\dms\to e^+e^-}
    \\\hline
        \scalaroperator{S} &
            2 s-8 m_e^2
    \\
        \scalaroperator{P} &
            2 s
    \\\hline\hline
        \text{Operator} &
        y_e^{-2}\coup^{-2}\supp^{4}\sum_{\text{spin}}
            \left|\mathcal M\right|^2_{\dms\bar\dms\to e^+e^-}
    \\\hline
        \scalaroperator{V} &
            -8\left(
                t-m_e^2
            \right)\left(
                s + t - m_e^2
            \right) + 16m_\dms^2\left(
                t-m_e^2
            \right) - 8m_\dms^4
    \\
        \scalaroperator{A} &
            -8t(s + t) + 16m_e^2t + 16m_\dms^2\left(
                t + m_e^2
            \right) - 8m_e^4 - 8m_\dms^4
    \\[0.1cm]\hline
    \end{array}
    \end{equation*}
    }
    \caption{Squared matrix elements for $\dms\bar\dms\to e^+e^-$ with $\dms$ a complex scalar, summed over final spin states. The operators are as defined in~\cref{tab:eft-scalar}. Note that the matrix elements for $\scalaroperator V$ and $\scalaroperator A$ vanish if $\dms$ is taken to be a real scalar. The matrix elements for scattering, $\dms e^-\to\dms e^-$, are obtained from these by the substitution $s\leftrightarrow t$.}
    \label{tab:scalar-squared-matrix-elements}
\end{table*}

\begin{table*}[htp]
    \centering
    {
    \setlength\Cwidth{7cm}
    \def\arraystretch{2}
    \begin{equation*}
    \begin{array}{
        |c|@{\hspace{1em}}l@{\hspace{1em}}|
    }
    \hline
        \text{Operator} &
        \coup^{-2}\supp^{2}\sigma(\dms\bar\dms\to e^+e^-)
    \\\hline
        \scalaroperator{S} &\displaystyle
            \frac{1}{8\pi s}\left(
                s-4 m_e^2
            \right)^{3/2}\left(
                s-4 m_\dms^2
            \right)^{-1/2}
    \\
        \scalaroperator{P} &\displaystyle
            \frac{1}{8\pi}\left(
                s-4 m_e^2
            \right)^{1/2}\left(
                s-4 m_\dms^2
            \right)^{-1/2}
    \\[0.1cm]\hline\hline
        \text{Operator} &
        y_e^{-2}\coup^{-2}\supp^{4}\sigma(\dms\bar\dms\to e^+e^-)
    \\\hline
        \scalaroperator{V} &\displaystyle
            \frac{1}{12\pi s}\left(
                s + 2 m_e^2
            \right)\left(
                s-4 m_e^2
            \right)^{1/2}\left(
                s-4 m_\dms^2
            \right)^{1/2}
    \\
        \scalaroperator{A} &\displaystyle
            \frac{1}{12\pi s}\left(
                s-4 m_e^2
            \right)^{3/2}\left(
                s-4 m_\dms^2
            \right)^{1/2}
    \\[0.1cm]\hline
    \end{array}
    \end{equation*}
    }
    \caption{Cross sections for $\dms\bar\dms\to e^+e^-$ for each effective operator in \cref{tab:eft-scalar}, summed over final spins. Note that the matrix elements for $\scalaroperator V$ and $\scalaroperator A$ vanish if $\dms$ is taken to be a real scalar.}
    \label{tab:scalar-annihilation-cross-sections}
\end{table*}

\begin{table*}[htp]
    {
    \setlength\Cwidth{7cm}
    \def\arraystretch{2.5}
    \begin{equation*}
    \begin{array}{|c|@{\hspace{1em}}l@{\hspace{1em}}|}
    \hline
        \text{Operator} &
        \coup^{-2}\supp^{2}\sigma(\dms e^-\to\dms e^-)
    \\\hline
        \scalaroperator{S} &\displaystyle
            \frac{1}{16\pi s^2}\left[
                s^2 + 6m_e^2s - 2m_\dms^2(s + m_e^2) + m_\dms^4 + m_e^4
            \right]
    \\
        \scalaroperator{P} &\displaystyle
            \frac{1}{16\pi s^2}\left[
                (s-m_e^2)^2 - 2m_\dms^2 (s+m_e^2) + m_\dms^4
            \right]
    \\[0.2cm]\hline\hline
        \text{Operator} &
        y_e^{-2}\coup^{-2}\supp^{4}\sigma(\dms e^-\to\dms e^-)
    \\\hline
        \scalaroperator{V} &\displaystyle
            \frac{1}{16\pi s}\left[
                s^2 + 2\left(
                    m_e^2 + m_\dms^2
                \right)s - \left(
                    m_e^2 - m_\dms^2
                \right)^2
            \right]
    \\
        \scalaroperator{A} &\displaystyle
            \frac{1}{16\pi s}\left[
                s^2 - 6m_e^2s + 2m_\dms^2\left(
                    s + m_e^2
                \right) - m_e^4 - m_\dms^4
            \right]
    \\[0.2cm]\hline
    \end{array}
    \end{equation*}
    }
    \caption{Cross sections for $\dms e^-\to \dms e^-$ for each effective operator in \cref{tab:eft-scalar}, averaged over initial spins and summed over final spins. Note that the matrix elements for $\scalaroperator V$ and $\scalaroperator A$ vanish if $\dms$ is taken to be a real scalar.}
    \label{tab:scalar-scattering-cross-sections}
\end{table*}

\begin{table*}[htp]
    {
    \setlength\Cwidth{7cm}
    \def\arraystretch{1.7}
    \begin{equation*}
    \begin{array}{|c|@{\hspace{1em}}l@{\hspace{1em}}|}
    \hline
        \text{Operator} &
        \coup^{-2}\supp^4\sum_{\text{spin}}
            \left|\mathcal M\right|^2_{\dmf\bar\dmf\to e^+e^-}
    \\\hline
        \fermionoperator{SS}
            & 4 \bigl(s-4 m_e^2\bigr)\bigl(s-4 m_\dmf^2\bigr)
    \\
        \fermionoperator{PS}
            & 4 s \bigl(s-4 m_e^2\bigr)
    \\
        \fermionoperator{SP}
            & 4 s \bigl(s-4 m_\dmf^2\bigr)
    \\
        \fermionoperator{PP}
            & 4 s^2
    \\
        \fermionoperator{VV}
            & 8\left(s+t\right)^2 + 8t^2 + 16m_+^4 - 32m_+^2t
    \\
        \fermionoperator{VA}
            & 8\left(s+t\right)^2 + 8t^2 + 16m_-^4 - 32m_+^2t - 32sm_e^2
    \\
        \fermionoperator{AV}
            & 8\left(s+t\right)^2 + 8t^2 + 16m_-^4 - 32m_+^2t - 32sm_\dmf^2
    \\
        \fermionoperator{AA}
            & 8\left(s+t\right)^2 + 8t^2 + 16m_+^2 - 32m_+^2t - 32m_+^2s +
                2\left(8m_em_\dmf\right)^2
    \\
        \fermionoperator{TT} &
            8(s+2t)^2 + 32m_+^4 - 16\left(s+4t\right)m_+^2 +
                \left(8m_em_\dmf\right)^2
    \\
        \fermionoperator{T\tilde T} &
            8(s+2t)^2 + 32m_-^4 - 16\left(s+4t\right)m_+^2
    \\[0.1cm]\hline
    \end{array}
    \end{equation*}
    }
    \caption{Squared matrix elements for $\dmf\bar\dmf\to e^+e^-$ with $\dmf$ a Dirac fermion, summed (not averaged) over initial and final spin states. The operators are as defined in~\cref{tab:eft-fermion}. Note that the matrix elements for $\fermionoperator{VV}$, $\fermionoperator{VA}$, $\fermionoperator{TT}$, and $\fermionoperator{T\tilde T}$ vanish if $\dmf$ is taken to be a Majorana fermion. For brevity, we define $m_\pm^2\equiv m_e^2\pm m_\dmf^2$. The matrix elements for scattering, $\dmf e^-\to\dmf e^-$, are obtained from these by the substitution $s\leftrightarrow t$.}
    \label{tab:dirac-squared-matrix-elements}
\end{table*}

\begin{table*}[htp]
    {
    \def\arraystretch{2.5}
    \begin{equation*}
    \begin{array}{|c|@{\hspace{1em}}l@{\hspace{1em}}|
                  |c|@{\hspace{1em}}l@{\hspace{1em}}|}
    \hline
        \text{Operator} &
        \coup^{-2}\supp^4\sigma(\dmf\bar\dmf\to e^+e^-)
        &
        \text{Operator} &
        \coup^{-2}\supp^4\sigma(\dmf\bar\dmf\to e^+e^-)
    \\\hline
        \fermionoperator{SS} &\displaystyle
            \frac{1}{16\pi}\frac{T_e^3T_\dmf}{s}
        &
        \fermionoperator{VA} &\displaystyle
            \frac{1}{12\pi}\frac{T_e^3}{sT_\dmf}\left(
                s+2m_\dmf^2
            \right)
    \\
        \fermionoperator{PS} &\displaystyle
            \frac{1}{16\pi}\frac{T_e^3}{T_\dmf}
        &
        \fermionoperator{AV} &\displaystyle
            \frac{1}{12\pi}\frac{T_eT_\dmf}{sT_e}\left(
                s+2m_e^2
            \right)
    \\
        \fermionoperator{SP} &\displaystyle
            \frac{1}{16\pi}T_eT_\dmf
        &
        \fermionoperator{AA} &\displaystyle
            \frac{1}{12\pi}\frac{T_e}{T_\dmf}\left[
                s^2 - 4\left(m_\dmf^2+m_e^2\right)s + 28m_\dmf^2m_e^2
            \right]
    \\
        \fermionoperator{PP} &\displaystyle
            \frac{1}{16\pi}\frac{sT_e}{T_\dmf}
        &
        \fermionoperator{TT} &\displaystyle
            \frac{1}{24\pi}\frac{T_e}{sT_\dmf}\left[
                \left(
                    s+2m_e^2
                \right)s + 2m_\dmf^2\left(
                    s+20m_e^2
                \right)
            \right]
    \\
        \fermionoperator{VV} &\displaystyle
            \frac{1}{12\pi}\frac{T_e}{T_\dmf}\left(
                s+2m_e^2
            \right)\left(
                s+2m_\dmf^2
            \right)
        &
        \fermionoperator{T\tilde T} &\displaystyle
            \frac{1}{24\pi}\frac{T_e}{sT_\dmf}\left[
                \left(
                    s+2m_e^2
                \right)s + 2m_\dmf^2\left(
                    s-16m_e^2
                \right)
            \right]
    \\[0.3cm]\hline
    \end{array}
    \end{equation*}
    }
    \caption{Cross sections for $\dmf\bar\dmf\to e^+e^-$ for each effective operator in \cref{tab:eft-fermion}, averaged over initial spins and summed over final spins. Note that the cross sections for $\fermionoperator{VV}$, $\fermionoperator{VA}$, $\fermionoperator{TT}$, and $\fermionoperator{T\tilde T}$ vanish if $\dmf$ is taken to be a Majorana fermion. For brevity, we define $T_i^2\equiv s-4m_i^2$.}
    \label{tab:dirac-annihilation-cross-sections}
\end{table*}

\begin{table*}[htp]
    \def\arraystretch{2.0}
    \begin{equation*}
    \begin{array}{|c|@{\hspace{1em}}l@{\hspace{1em}}|}
    \hline
        \text{Operator} &
        48\pi s^3\coup^{-2}\supp^4\sigma(\dmf e^-\to \dmf e^-)
    \\\hline
        \fermionoperator{SS} &\displaystyle
            s^4 + 2m_+^2s^3 + 2s^2\left(3m_e^4-14m_e^2m_\dmf^2+3m_\dmf^4\right)
            + 2m_-^4m_+^2s + m_-^8
    \\
        \fermionoperator{PS} &\displaystyle
            \left(
                s^2 + 4sm_e^2 + m_e^4 + m_\dmf^4 - 2m_\dmf^2s_e^+
            \right)\left(
                {s_e^-}^2 - 2m_\dmf^2s_e^+ + m_\dmf^4
            \right)
    \\
        \fermionoperator{SP} &\displaystyle
            \left[
                m_\dmf^2\left(
                    4s - 2m_e^2 + m_\dmf^2
                \right) + {s_e^-}^2
            \right]\left(
                {s_e^-}^2 - 2m_\dmf^2s_e^+ + m_\dmf^4
            \right)
    \\
        \fermionoperator{PP} &\displaystyle
            \left(
                {s_e^-}^2 - 2m_\dmf^2s_e^+ + m_\dmf^4
            \right)^2
    \\
        \fermionoperator{VV} &\displaystyle
            2s^2\left(
                4s^2 - 10m_+^2s
                + 9m_e^4 + 22m_e^2m_\dmf^2 + 9m_\dmf^4
            \right) - 8m_+^2m_-^4s + 2m_-^8
    \\
        \fermionoperator{VA} &\displaystyle
            2\left(
                {s_e^-}^2 - 2m_\dmf^2s_e^+ + m_\dmf^4
            \right) \left[
                (s + s_e^+)^2 - 2m_\dmf^2s_e^+ + m_\dmf^4
            \right]
    \\
        \fermionoperator{AV} &\displaystyle
            2\left[
                2s\left(
                    2s - m_e^2 + 2m_\dmf^2
                \right) + m_-^4
            \right]\left(
                {s_e^-}^2 - 2m_\dmf^2s_e^+ + m_\dmf^4
            \right)
    \\
        \fermionoperator{AA} &\displaystyle
            2s^2\left(
                4s^2 - 4m_+^2s
                - 3m_e^4 + 46m_e^2m_\dmf^2 - 3m_\dmf^4
            \right) + 4m_+^2m_-^4s + 2m_-^8
    \\
        \fermionoperator{TT} &\displaystyle
            2s^2\left(
                7s^2 - 13m_+^2s
                + 6m_e^4 + 52m_e^2m_\dmf^2 + 6m_\dmf^4
            \right) - 2m_+^2m_-^4 + 2m_-^8
    \\
        \fermionoperator{T\bar{T}} &\displaystyle
            2\left[
                s\left(
                    m_+^2 + 7s
                \right) + m_-^4
            \right]\left(
                {s_e^-}^2 - 2m_\dmf^2s_e^+ + m_\dmf^4
            \right)
    \\[0.3cm]\hline
    \end{array}
    \end{equation*}
    \caption{Cross sections for $\dmf e^-\to \dmf e^-$ for each effective operator in \cref{tab:eft-fermion}, averaged over initial spins and summed over final spins. Note that the cross sections for $\fermionoperator{VV}$, $\fermionoperator{VA}$, $\fermionoperator{TT}$, and $\fermionoperator{T\tilde T}$ vanish if $\dmf$ is taken to be a Majorana fermion.For brevity, we define $m_\pm^2\equiv m_e^2\pm m_\dmf^2$ and $s_i^\pm\equiv s\pm m_i^2$.}
    \label{tab:dirac-scattering-cross-sections}
\end{table*}

\end{document}